\documentclass[aps,prl,twocolumn,amsmath,amssymb]{revtex4-1}

\usepackage{graphicx}
\usepackage{color}
\usepackage{epstopdf}
\usepackage{amsmath}
\usepackage{bm}
\usepackage{hyperref}
\usepackage{cleveref}
\usepackage{tcolorbox}
\usepackage{braket}
\usepackage[export]{adjustbox}
\usepackage[hang]{subfigure}
\usepackage{makecell}
\usepackage{float}
\usepackage{booktabs}
\usepackage[section]{placeins}
\usepackage{verbatim}

\begin{document}
	
	\title{The Solid-state Physics of Rydberg-dressed Bosonic Mixtures}
	
	\author{Yi-Ming Duan$^{1}$}
	\thanks{These authors contribute equally to this work.}
	
    \author{Liang-Jun He$^{1}$}
    \thanks{These authors contribute equally to this work.}
    
    \author{Fabian Maucher$^{2}$}
    
    \author{Yong-Chang Zhang$^{1}$}
    \email{zhangyc@xjtu.edu.cn}
    
	\affiliation{$^1$MOE Key Laboratory for Nonequilibrium Synthesis and Modulation of Condensed Matter, Shaanxi Key Laboratory of Quantum Information and Quantum Optoelectronic Devices, School of Physics, Xi'an Jiaotong University, Xi'an 710049, People's Republic of China\\
	$^2$Faculty of Mechanical Engineering; Department of Precision and Microsystems Engineering, Delft University of Technology, 2628 CD, Delft, The Netherlands}
		
\begin{abstract}
	We explore phases of two-component Rydberg-dressed Bose-Einstein condensates in three spatial dimensions. The competition between the effective ranges of inter- and intra-component soft-core interactions leads to a rich variety of ground states. These include states resembling ionic compounds with face-centered cubic or simple cubic lattice structure. Upon increasing the scattering length, the dimensionality of the symmetry-breaking is lower due to the suppression of large densities, leading to segregated planar or tubular density profiles. We also show that these states are not only stable ground states, but can also emerge dynamically upon time evolution.
\end{abstract}

\maketitle

\emph{Introduction---}Ultracold quantum gases represent a unique platform for exploring many-body physics and fundamental quantum behavior due to the tunability and control of interatomic interactions~\cite{Bloch2008RMP,Saffman:RMP:2010,Langen2015Review,Gross2017Science}. Two prominent examples for long-range interacting systems are dipolar Bose-Einstein condensates (dBECs) and Rydberg-dressed Bose-Einstein condensates (RdBECs). The pioneering experiment~\cite{Kadau2016,Schmidt2016Nature} showed that dBECs can be stabilized against collapse through quantum fluctuations, which paved the way for the observation of crystallization and supersolidity~\cite{Tanzi2019PRL,Chomaz2019PRX,Bottcher2019PRX,Modugno:2021:Science} in this system. Whereas collapse can also be arrested by employing a suitable trapping~\cite{Koch2008,Pfau2009review,Chomaz2023review}, crystallization is inhibited by such a confinement. Since dBECs feature an anisotropic interaction where it is attractive along the polarization direction, external confinement by either a cigar- or pancake-shaped trap is necessary to promote one-dimensional~\cite{Tanzi2019PRL,Chomaz2019PRX,Bottcher2019PRX} or two-dimensional (2D)~\cite{norcia2021two} symmetry-breaking, respectively, and three-dimensional (3D) symmetry-breaking is not possible. 

In contrast, RdBECs permit the realization of symmetry-breaking in all three spatial directions. RdBECs are realized by off-resonantly coupling the atomic ground state and the Rydberg-excited state via optical fields. This leads to an admixing of Rydberg-character to the atomic ground state, thus two distant atoms feature an effective van der Waals interaction whilst two close atoms are blocked from being doubly excited, inducing a soft-core interaction~\cite{PhysRevLett.85.2208,PhysRevA.82.033412,PhysRevLett.105.160404,Pfau:NJP:2014}. Through suitable confinement, this platform also allows the exploration of supersolidity in 2D systems~\cite{Henkel2012PRL,PhysRevLett.104.223002,Cinti2014NC}. If there is no confinement, it has been predicted that RdBECs can evolve into 3D face-centered cubic (FCC) crystals~\cite{Henkel2010PRL,Ancilotto2013PRA}.

\begin{figure}[!ht]
	\centering
	\includegraphics[width=\columnwidth]{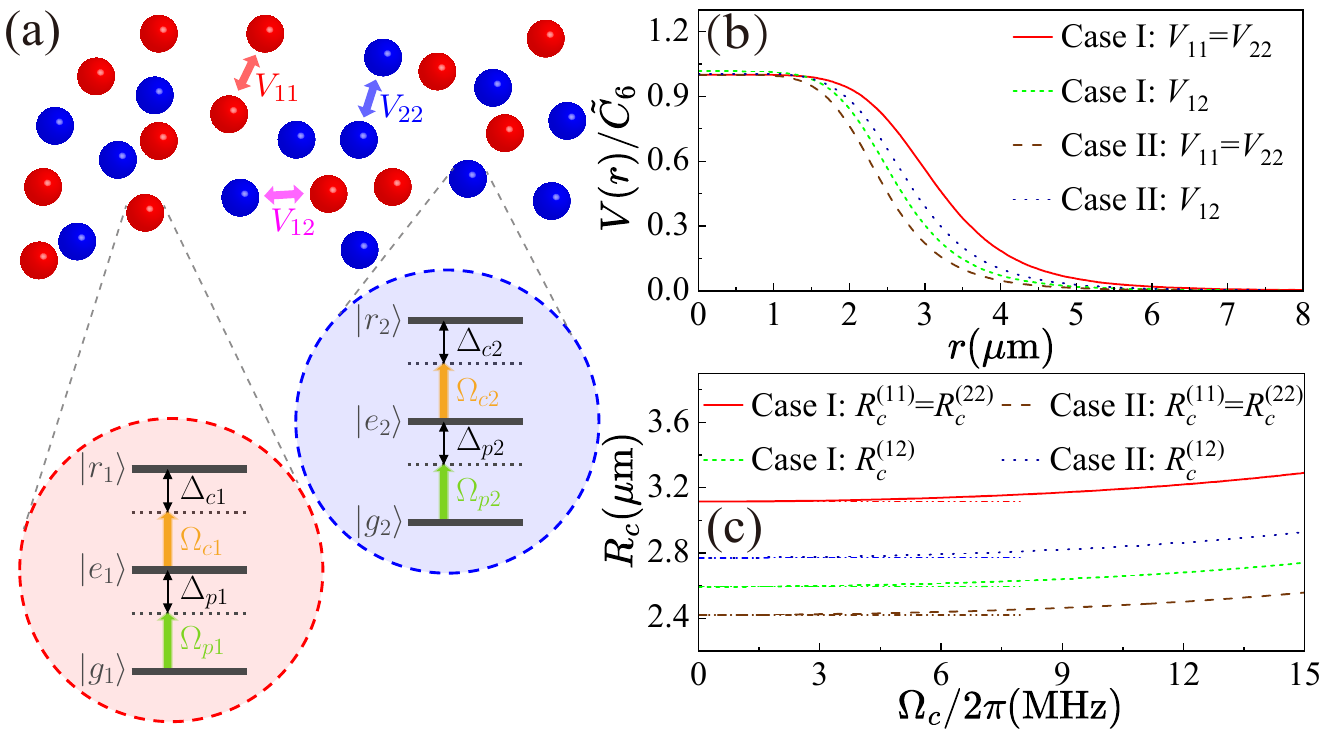}
	\caption{(a) Illustration of two-component RdBECs. The two hyperfine ground states $|g_1\rangle$ (component 1) and $|g_2\rangle$ (component 2) are coupled to the two excited states $|e_1\rangle$ and $|e_2\rangle$, respectively, by external optical fields with Rabi frequencies $\Omega_{p1}$, $\Omega_{p2}$ and detunings $\Delta_{p1}$, $\Delta_{p2}$. 
    A further optical field with Rabi frequency $\Omega_{c1}$ ($\Omega_{c2}$) off-resonantly drives the atomic transition $|e_1\rangle \leftrightarrow |r_1\rangle$ ($|e_2 \rangle \leftrightarrow |r_2\rangle$) with the detuning $\Delta_{c1}$ ($\Delta_{c2}$). Here $|r_1\rangle$ and $|r_2\rangle$ are two Rydberg states. $V_{11,22}(\mathbf{r})$ and $V_{12}(\mathbf{r})$ represent the respective intra- and inter-component soft-core potentials between ground-state atoms induced by Rydberg dressing. (b) The soft-core potentials vary with atomic separation for the situations of a smaller (Case I) and a larger (Case II) inter-component blockade radius $R^{(12)}_c$ in comparison with the intra-component counterpart $R^{(11)}_c=R^{(22)}_c$ (see the relevant parameters in the Supplemental Material~\cite{supp}). (c) The variation of inter- and intra-component blockade radii with respect to the Rabi frequencies $\Omega_c=\Omega_{c1}=\Omega_{c2}$, where the horizontal dashed lines refer to the corresponding values in the large detuning limit~\cite{Henkel2010PRL,Hsueh2013PRA}.}
	\label{fig1}
\end{figure}

\nocite{Nikola2017ARC,Nikola2021ARC,HenkelPhD,PhysRevLett.95.248301}

In a spatially multidimensional system~\cite{Copar2019PRX,Biagioni2022PRX}, adding a second species~\cite{Pu1998PRL1,Coen2001PRL,Schneider2014PRC,Igor2015NJP} allows access to even more complexity and intricate behaviors~\cite{Pu1998PRL2,Shoji2007PRE,PhysRevLett.133.103401}. For example, in 2D systems, a range of unusual phases such as N{\'e}el, ladder and Peierls lattices~\cite{Hsueh2013PRA} as well as chiral states~\cite{Han2018PRL} were predicted in binary RdBECs. The two-component studies were mostly focused on 2D RdBECs. 
\textcolor{black}{Therefore, it appears promising to explore what types of Bravais lattices~\cite{Kittel2004,JAP2016,RatioRule,Li:Chem:2023} and supersolidity featuring corresponding spatial symmetries emerge in 3D systems}.


In this work, we address that by considering a 3D two-component RdBEC subject to both inter- and intra-component soft-core interactions as depicted in Fig.~\ref{fig1}. We show that this system supports a rich variety of ground states beyond what has been predicted and is unique to 3D long-range interacting multi-species BECs. 

\emph{Theoretical model---}As illustrated in Fig.~\ref{fig1}, the Rydberg atoms are prepared in the two hyperfine ground states $|g_1\rangle$ (component 1) and $|g_2\rangle$ (component 2)~\cite{Hsueh2013PRA}. These states are off-resonantly coupled to the highly excited Rydberg states $|r_1\rangle$ and $|r_2\rangle$ through the intermediate excited states $|e_1\rangle$ and $|e_2\rangle$ by the corresponding driving fields, respectively. In the situation of off-resonant coupling with large detunings, the populations of excited sates are negligibly small, and the ground-state atoms interact via an effective long-range soft-core potential $V_{\alpha\beta}(\mathbf{r})=\frac{\tilde{C}_6^{(\alpha\beta)}}{1+({r}/{R_c^{(\alpha\beta)}})^6}$ $(\alpha,\beta=1,2)$. 
We find that the Rydberg blockade radius 
$R_c^{(\alpha\beta)}=( \frac{C_6^{(\alpha\beta)}}{ {\Omega^2_{c\alpha}/{\Delta_{p\alpha}}} + {\Omega^2_{c\beta}/{\Delta_{p\beta}}-(\Delta_{c\alpha}+\Delta_{p\alpha}+\Delta_{c\beta}+\Delta_{p\beta})}}) ^{1/6}$ and the effective interaction coefficient $\tilde{C}_6^{(\alpha\beta)}=\frac{\hbar\Omega^2_{c\alpha}\Omega^2_{c\beta}\Omega^2_{p\alpha}\Omega^2_{p\beta}  }{\Delta_{p\alpha}\Delta_{p\beta}}\times \frac{  \Delta_{p\beta}(\Omega^2_{c\alpha}-\Delta_{p\alpha}(\Delta_{c\alpha}+\Delta_{p\alpha})) + \Delta_{p\alpha}(\Omega^2_{c\beta}-\Delta_{p\beta}(\Delta_{c\beta}+\Delta_{p\beta})) }{(\Omega^2_{c\alpha}-\Delta_{p\alpha}(\Delta_{c\alpha}+\Delta_{p\alpha}))^2(\Omega^2_{c\beta}-\Delta_{p\beta}(\Delta_{c\beta}+\Delta_{p\beta}))^2}$ can be tuned well through the driving fields~\cite{supp}. 
For sufficiently large detunings, they reduce to expressions that have previously been found for RdBECs~\cite{Henkel2010PRL,Hsueh2013PRA,Han2018PRL}. The dynamics of this binary condensate is governed by the following two coupled Gross-Pitaevskii equations (GPE):
\begin{equation}
	\begin{split}
		i\hbar \frac{\partial}{\partial t}\Psi_\alpha (\mathbf{r})&=-\frac{\hbar^{2} \nabla^{2}}{2 M}{\Psi_\alpha}(\mathbf{r})
		+\sum_{\beta=1,2}\big[ g_{\alpha\beta}\left|\Psi_\beta (\mathbf{r}) \right|^2\\
		&+\int \mathrm{d}^{3} \mathbf{r}'
		V_{\alpha\beta}\left(\mathbf{r}-\mathbf{r}'\right)\left|\Psi_\beta\left(\mathbf{r}'\right)\right|^2\big]\Psi_\alpha(\mathbf{r}).
	\end{split}
	\label{eq:1}
\end{equation}
Here, $\Psi_\alpha (\mathbf{r})$ denotes the wave function of the $\alpha$-th component and is normalized to the particle number $N_\alpha=\int |\Psi_\alpha(\mathbf{r})|^2 \mathrm{d}^{3} \mathbf{r}$, and $g_{\alpha\beta}={4\pi\hbar^2 a_s^{(\alpha\beta)}}/{M}$ represents the contact interaction strength with $M$ being the atomic mass and $a^{(\alpha\beta)}_s$ the $s$-wave scattering length.

In the following, we focus on the symmetric configuration, where $g\equiv g_{11}=g_{22}=g_{12}$, $\tilde{C}_6\equiv \tilde{C}_6^{(11)}=\tilde{C}_6^{(22)}=\tilde{C}_6^{(12)}$, $R_c\equiv  R^{(11)}_c=R^{(22)}_c$, and $N_1=N_2$. However, the parameter space is expanded by the fact that $R^{(12)}_c$ is not necessarily equal to $R_c$ as considered in~\cite{Hsueh2013PRA,Han2018PRL} [cf. Fig.~\ref{fig1}(b,c)]. This imbalance between the intra- and inter-component blockade radii can be achieved by selecting the Rydberg states adequately and carefully tuning the driving fields~\cite{Igor2015NJP,Qian2015PRA,Igor2016NJP,Igor2016PRA,Igor2019PRE,supp}. 

\begin{figure*}[!t]
	\centering
	\includegraphics[width=\linewidth]{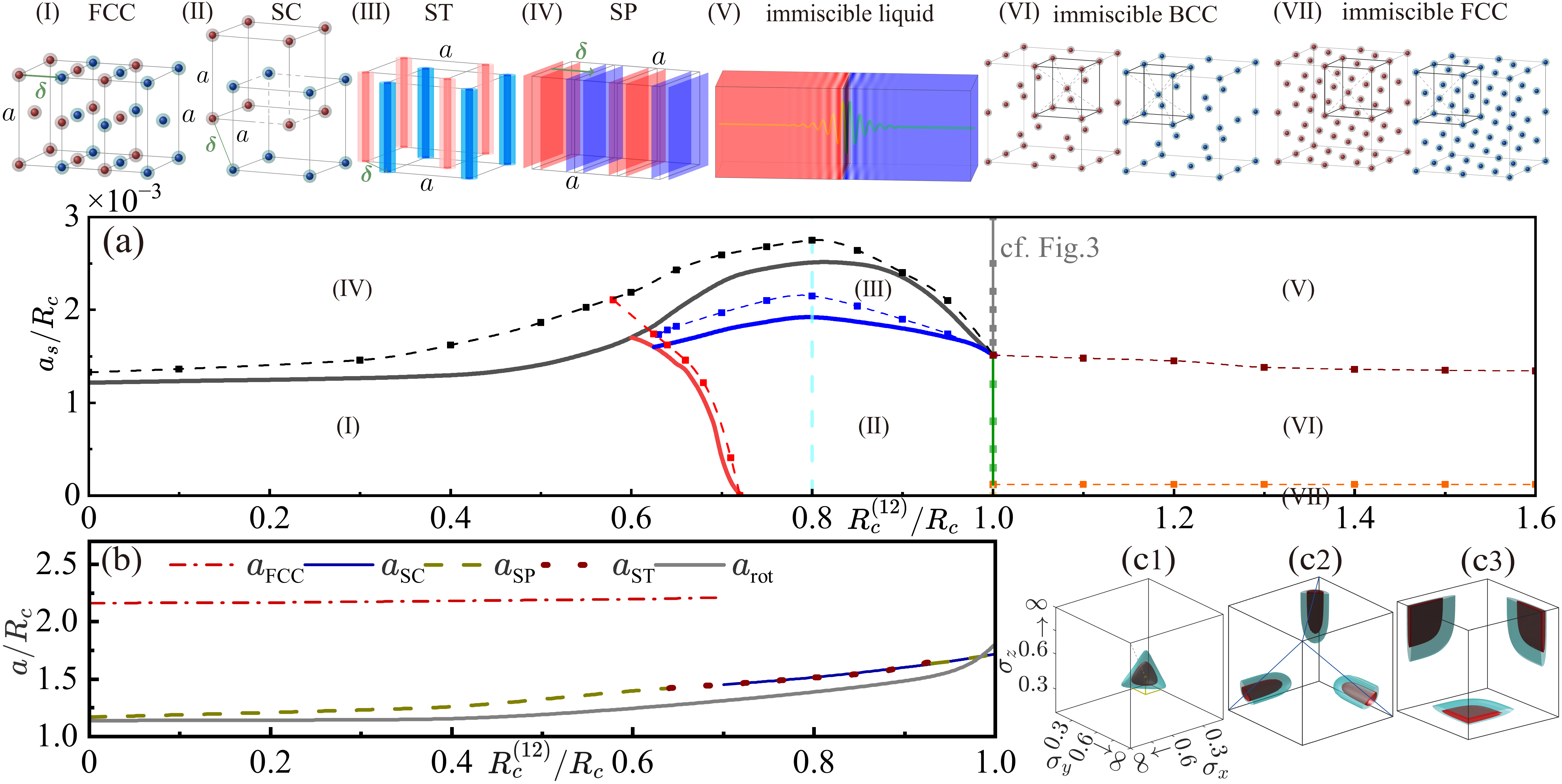}
	\caption{(a) Ground-state phase diagram of the 3D binary RdBEC as a function of the contact interaction $a_s$ and the inter-component blockade radius $R^{(12)}_c$. The phase boundaries obtained variationally and numerically are presented by the solid lines and by the points on the dashed line, respectively. Here we have fixed the parameters $\tilde{C}_6 =0.15 \hbar^2/(MR^2_c)$ and $\rho R^3_c=1.5\times 10^{3}$. (b) The lattice constant $a$ of each state as well as the roton-determined length scale $a_{\rm rot}=2\pi/k_{\rm rot}$ varies with $R^{(12)}_c/R_c$. Here, the $s$-wave scattering lengths have been fixed at $a_s/R_c=0.5\times  10^{-3}$ and $2.3\times 10^{-3}$ for the FCC state and the other three states, respectively, although the lattice constant of each state actually does not vary much with $a_s$. (c1$\sim$c3) exhibit the corresponding variational energy landscapes along the cyan dashed line in (a) at $a_{s} /R_c=1.0\times 10^{-3}$, $2.3\times 10^{-3}$, and $3.0\times 10^{-3}$, respectively.}
	\label{fig2}
\end{figure*}

\emph{Ground-state phase diagram---}To avoid finite-size effects, we explore the ground states of this system in the thermodynamic limit, i.e., we consider an infinite particle number ($N_\alpha \rightarrow \infty$) and an infinite spatial volume ($\mathcal{V}\rightarrow \infty$) whilst the average density remains finite $\rho=N_1/\mathcal{V}=N_2/\mathcal{V}$. Numerically, that is realized by using periodic boundary conditions. We employ both imaginary-time evolution and Gaussian variational analysis to identify the ground states of the GPE~\eqref{eq:1}. Further details are provided in the Supplemental Material~\cite{supp}. The resulting phase diagram is shown in Fig.~\ref{fig2}, 
where the spatial and energy units are rescaled by $R_c$ and $\hbar^2/(MR^2_c)$, respectively.

In the case of a small inter-component Rydberg blockade radius and weak contact interactions, where the intra-component long-range interactions dominate [cf. the lower-left area in Fig.~\ref{fig2}(a)], the ground states of both components exhibit a typical FCC profile that is translated with respect to each other by $\delta=a/2$. Here, $a$ represents the lattice constant. The total density distribution -- i.e., neglecting the fact that there are two species -- forms a simple cubic (SC) crystal with a lattice constant of $\delta$. Upon increasing the $s$-wave scattering length, the local repulsion renders high-density regions energetically expensive. Therefore, it forces the atoms to move away from each lattice site in order to minimize the energy related to contact interaction. 
In other words, for sufficiently strong contact interactions, the large local interaction strength avoids higher-dimensional symmetry-breaking due to the suppression of high densities, thereby leading to a ``smearing out" of density modulation and, in principle, one could ultimately expect an overall unmodulated density as ground state when increasing the scattering length sufficiently. 

However, in contrast to 3D single-component and 2D two-component systems with equal blockade radii, where crystal phases typically melt into a uniform profile, the ground state of our system undergoes a surprising transition to a state composed of alternating segregated planar (SP) density distributions if the contact interaction exceeds a certain threshold. Thus, the continuous translational symmetry one could expect from the single-component analogue is broken due to the imbalance between the inter- and intra-component blockade radii. 

The emergence of these SP states can be understood from the Bogoliubov excitation spectrum of the miscible flat mixtures, which features two branches as follows:
\begin{equation}
    \omega_\pm(k)=k\sqrt{k^2/4+\rho\left[g+\mathcal{U}(k)\right]\pm \rho \left[g+\mathcal{U}^{(12)}(k)\right]}.
    \label{eq:2x}
\end{equation}
Here, $\mathcal{U}(k)$ and $\mathcal{U}^{(12)}(k)$ are the Fourier transform of the intra- and inter-component soft-core potentials, respectively. 
The two branches represent in- and out-of-phase excitations of the two species with modulation-direction ${\bf k}$ and modulation frequency $2\pi/k$. We find that the out-of-phase excitations are energetically favorable, thereby offering a useful elucidation as to why the SP phase emerges. Furthermore, the roton-minimum $k_{\rm rot}$ of the lower branch $\omega_-(k)$ provides an estimate for the spatial periodicity of the density modulation. Fig.~\ref{fig2}(b) exhibits this length scale $a_{\rm rot}=2\pi/k_{\rm rot}$, showing reasonable agreement with the lattice constant of the SP phase.

Furthermore, Fig.~\ref{fig2}(b) presents the dependencies of the lattice constants for the other possible lattices as a function of $R^{(12)}_c/R_c$.
Upon enlarging $R^{(12)}_c$, the range of the inter-component repulsion increases, prompting the components to favor a larger spatial inter-component separation. We note a sudden drop in lattice constant $a$ from the FCC structure to the SC or segregated tubular (ST) lattice. However, whilst the lattice constant decreases, the displacement $\delta$ between the structures increases from $\delta_{\rm FCC}=a/2$ to $\delta_{\rm ST}=a/\sqrt{2}$ and $\delta_{\rm SC}=\sqrt{3}a/2$. With respect to the SC state, if we were to neglect the fact that there are two components, the entire density distribution would feature a body-centered cubic (BCC) lattice. 

As in the previous scenario with smaller $R^{(12)}_c$, the SP lattice emerges as the ground state at sufficiently strong contact interactions. However, at intermediate strengths of the contact interaction, i.e., the area between the SC and SP states in Fig.~\ref{fig2}(a), a surprising ST state becomes the ground state. This density profile is reminiscent of the filaments found in confined dBECs~\cite{Blakie2018PRL} and in biological systems~\cite{Park2019Bio}. It is a comparably rare distribution that is not often seen in other systems. The phase transition from the SC state to the ST state is of the second order. Upon further increasing the contact interaction, again a second-order phase transition occurs towards the SP state. 
These SP and ST states resemble the so-called lasagne and spaghetti phases in neutron stars~\cite{Schneider2013PRC,Horowitz2015PRL,Caplan2017RMP,Caplan2018PRL}. 
\textcolor{black}{In a different context, the mimicking nature of supersolids in BECs and neutron stars yielded an explanation to glitches in rotation frequency~\cite{PhysRevLett.131.223401}.} 
\textcolor{black}{In order to highlight the lattice structures, Fig.~\ref{fig2} presents the isosurfaces at a large density. In reality, there is overlap (i.e., superfluid background) between neighboring droplets, tubes, or planes, which is a characteristic signature of supersolids~\cite{Tanzi2019PRL,Chomaz2019PRX,Bottcher2019PRX,supp,PhysRevLett.25.1543,Leggett1998}.}

To gain a better understanding of these states, we also performed a Gaussian variational analysis by approximating the density profile of each site with a Gaussian distribution~\cite{santi2018freezing,santi2020a,Prestipino_2019,Blakie2020var}. In this approach, the overall density profile of the SC state can be described by
\begin{equation}
	\begin{split}
	\rho_{1}=&\frac{1}{\mathcal{N}}\sum_{l,m,n}e^{-\left[\frac{\left(x+l a\right)^2}{\sigma_x^2}+\frac{\left(y+m a\right)^2}{\sigma_y^2}+\frac{\left(z+n a\right)^2}{\sigma_z^2}\right]}, \\
	\rho_{2}=&\frac{1}{\mathcal{N}}\sum_{l,m,n}e^{-\left[\frac{\left(x+l a+\delta_x\right)^2}{\sigma_x^2}+\frac{\left(y+m a+\delta_y\right)^2}{\sigma_y^2}+\frac{\left(z+n a+\delta_z\right)^2}{\sigma_z^2}\right]},
	\end{split}
	\label{eq:2}
\end{equation}
where $\mathcal{N}$ is the normalization coefficient, $l,m$ and $n$ are integers, $(\delta_x,\delta_y,\delta_z)$ denotes the misalignment vector between the two species, and $\sigma_{x,y,z}$ represents the width along the corresponding direction~\cite{supp}. Using these trial profiles, we obtain the total energy of the system as a function of the variational parameters $\sigma_{x,y,z}$ and $a$. By minimizing the energy, one can readily determine the ground states and gain a more complete understanding of the respective energy landscape without employing heavy numerics. Figs.~\ref{fig2}(c1$\sim$c3) display this energy landscape varying with $\sigma_{x,y,z}$ at the optimal lattice constant $a$. We find that there is a single minimum at $a_s/R_c=1.0\times 10^{-3}$ [see Fig.~\ref{fig2}(c1)], where $\sigma_{x}$, $\sigma_{y}$, and $\sigma_{z}$ are equal and finite, indicating a typical SC lattice profile for each component. As the contact interaction increases, the energy landscape undergoes a significant change, accompanied by the emergence of two additional minima as illustrated in Fig.~\ref{fig2}(c2). Hence, there are three degenerate minima at $(\sigma_x=\sigma_y,\ \sigma_z\rightarrow\infty)$, $(\sigma_x=\sigma_z,\ \sigma_y\rightarrow\infty)$, and $(\sigma_y=\sigma_z,\ \sigma_x\rightarrow\infty)$ for $a_s/R_c=2.3\times 10^{-3}$. The divergent width at these minima implies an infinitely long continuous distribution along the corresponding direction, characteristic for the ST state. Furthermore, at $a_s/R_c=3.0\times 10^{-3}$ [see Fig.~\ref{fig2}(c3)], only one width remains finite while the other two diverge, corresponding to the profile of the SP state. Additionally, this Gaussian variational ansatz can be extended to describe the FCC state as well as the BCC state. It permits to variationally determine the transition points between different phases [see the solid lines in Fig.~\ref{fig2}(a)], presenting a reasonable agreement with the accurate full GPE simulation results.

As $R^{(12)}_c$ increases further to the regime of $R^{(12)}_c/R_c >1$, the system exhibits phase-separation behaviour. That is, the two species are no longer able to merge into each other due to the strong inter-component repulsion. Instead, a completely immiscible density distribution is preferred. Within each component's domain, the system reduces to single-component physics, which can either feature crystalline lattices at small $a_s$ or segregated transversely flat states at large $a_s$. However, a noticeable evanescent density oscillation persists at the interface between the two species, induced by the inter-component long-range interaction. This modulation gradually diminishes as it moves into the respective domain of each component (cf. the yellow and green curves in the distribution of the immiscible liquid in Fig.~\ref{fig2}). 

\begin{figure}[!t]
 	\centering
 	\includegraphics[width=\columnwidth]{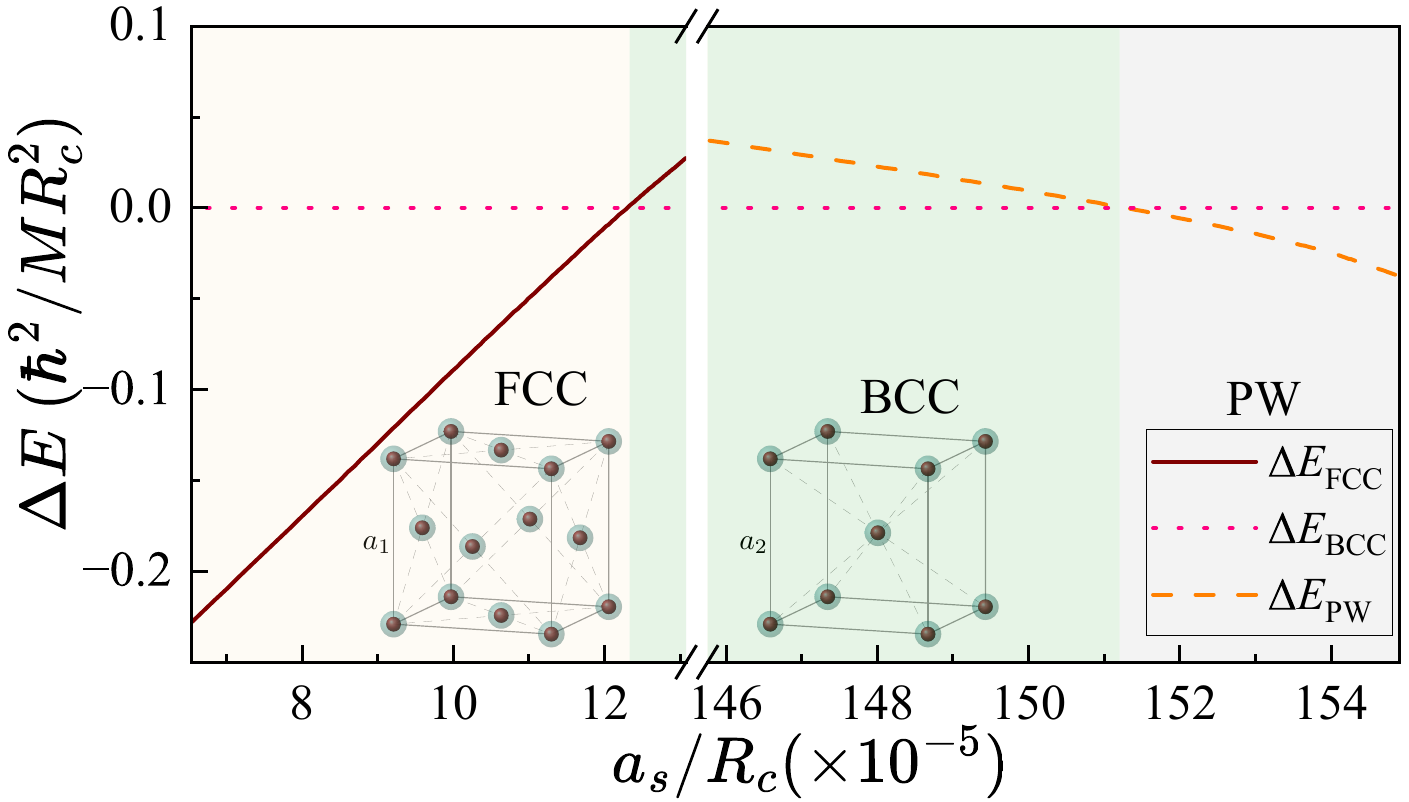}
 	\caption{The single-particle energy of each state with respect to the BCC state, i.e., $\Delta E_{\gamma}=E_{\gamma}-E_{\rm BCC}$ ($\gamma={\rm FCC,\ BCC,\ PW}$) varies with the contact interaction when $R^{(12)}_c/R_c =1$. Apart from that we used the same relevant parameters as in Fig.~\ref{fig2}.}
 	\label{fig3}
\end{figure}

At the specific point where $R^{(12)}_c/R_c =1$, the entire system reduces to a single-component RdBEC in the situation of strong interactions, as the energy is predominately governed by the total density distribution $|\Psi_1|^2+|\Psi_2|^2$. Fig.~\ref{fig3} shows the ground-state energy shift with respect to the BCC state as a function of the contact interaction. At weak contact interaction, the ground state features a conventional FCC structure with a lattice constant of $a_{1}=2.18R_c$. This is consistent with the result observed in the 3D single-component system with soft-core interactions only~\cite{Ancilotto2013PRA}. As the contact interaction exceeds the critical point ($a_s/R_c\sim 1.2\times 10^{-4}$), the RdBEC undergoes a first-order phase transition from FCC to BCC states, with a lattice constant of $a_{2}=1.69 R_c$. If $a_s$ is further increased beyond a threshold ($a_s/R_c\sim 1.51\times 10^{-3}$), the ground state eventually transitions into an unmodulated plane wave (PW) state due to the strong local repulsion. Furthermore, as shown in Fig.~\ref{fig2}(a), this transition point $a_s/R_c\sim 1.51\times 10^{-3}$ serves as a multi-critical point, where the previously discussed SC, ST, SP, and phase-separated states all terminate.

\begin{figure}[!t]
\centering
\includegraphics[width=\columnwidth]{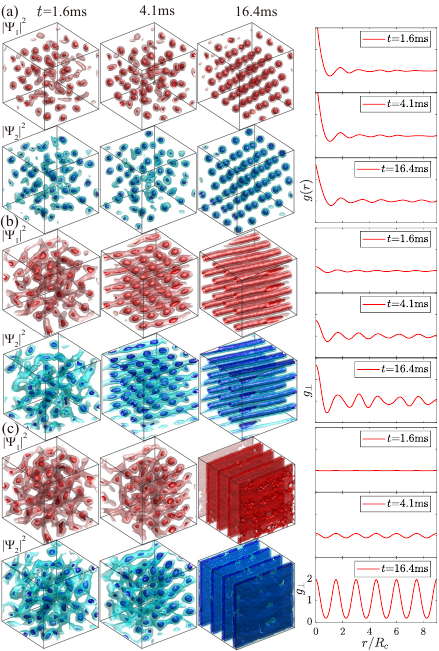}
\caption{The snapshots of dynamically evolving binary RdBECs at different times for $R_c^{(12)}/R_c=0.95$. The instantaneous density correlation function $g(r)$~\cite{supp} is depicted in the right column. Here, the intra-component blockade radius has been fixed at $R_c=4.5\times 10^4 a_0$ with $a_0$ being the Bohr radius, and the corresponding $s$-wave scattering lengths in (a$\sim$c) are $a_{s}/a_0 =40$, 80, and 120, respectively, while the other parameters are same as those in Fig.~\ref{fig2}.}
\label{fig4}
\end{figure}

\emph{Dynamical evolution---}To explore whether such states can be experimentally accessed, we study real-time evolution of this binary RdBEC. In the simulation, we initialize the condensate in a homogeneous unmodulated state with random noise and then propagate the GPE~\eqref{eq:1} in real time, using the fixed parameters $R_c=4.5\times 10^4 a_0$ with $a_0$ being the Bohr radius, $R_c^{(12)}/R_c=0.95$, $\rho R^3_c=1.5\times 10^{3}$, and $\tilde{C}_6=0.15\hbar^2/(MR_c^2)$ with $M$ the atomic mass of cesium. As shown in Fig.~\ref{fig4}, the system undergoes the dynamical formation of various structures depending on the strength of contact interaction over a time span of $17{\rm ms}$. For the case of weak contact interactions (e.g., $a_s/a_0=40$), the two species dynamically self-organize into SC structures around $t\sim 4.1 {\rm ms}$, eventually reaching equilibrium at $t\sim 16.4 {\rm ms}$. Moreover, visible characteristics of the ST and SP states emerge as time progresses in the stronger interaction cases of $a_s/a_0=80$ and $120$, respectively. 
Thereby the various profiles can be dynamically accessed and resemble the respective ground states. 

\emph{Conclusion---} In summary, we have demonstrated that 3D binary RdBECs subjected to contact and soft-core interactions support a rich variety of ground states. This zoo includes states resembling lattice structures of ionic compounds where each of the components crystallizes in FCC or SC structure that cannot be realized in a single component. A lattice structure that is known as ``tetrahedral hole"~\cite{RatioRule} can also be found as a metastable crystal for small $R_c^{(12)}$ (not shown). Furthermore, it also hosts immiscible fluids, immiscible crystals, and unusual states with SP or ST density profiles, \textcolor{black}{offering a promising avenue to explore interesting physics, e.g., in neutron stars~\cite{Schneider2013PRC,Horowitz2015PRL,Caplan2017RMP,Caplan2018PRL,PhysRevLett.131.223401}}.  
In the limit where the two-component condensate acts like a single-component system, apart from the FCC structure~\cite{Henkel2010PRL,Ancilotto2013PRA} we also report a new BCC profile. These can be made possible by tuning the ratio between the inter- and intra-component Rydberg blockade radii and can be dynamically accessed. Although it remains challenging to achieve long-lived RdBECs due to dissipation~\cite{PhysRevLett.116.113001,PhysRevA.105.013109}, 
recent experimental achievements feature significant steps towards stable RdBECs with a comparable long lifetime~\cite{PhysRevX.11.021036,PhysRevLett.120.183401, PhysRevLett.131.063401,Bloch:arxiv:2024}.

This platform offers further degrees of freedom beyond the tunable ratio between the inter- and intra-component blockade radii. For example, considering an unequal particle number, interaction strength, and intra-component blockade radii in the two components already expand possible parameter space significantly.  
Beyond that, one can also couple to $p$- or $d$-orbital Rydberg states~\cite{Maucher:PRL:2011} or adjust the polarization of the driving fields~\cite{PhysRevLett.130.243001} for the dressing, yielding anisotropic soft-core potentials, \textcolor{black}{which expands possible high-dimensional crystal and supersolidity structures even further}.

\emph{Acknowledgement---}We are grateful to Thomas Pohl for helpful discussions. This work was supported by the National Nature Science Foundation of China (Grant No.: 12104359), National Key Research and Development Program of China (Grant No.: 2021YFA1401700), Shaanxi Academy of Fundamental Sciences (Mathematics, Physics) (Grant No.: 22JSY036), and the Innovation Program for Quantum Science and Technology (2024ZD0300600). Y.C.Z. acknowledges the support of Xiaomi Young Talents program, Xi'an Jiaotong University through the ``Young Top Talents Support Plan" and Basic Research Funding as well as the High-performance Computing Platform of Xi'an Jiaotong University for the computing facilities. 

\bibliography{mybib}

\end{document}


\title{Supplemental Material for ``The Solid-state Physics of Rydberg-dressed Bosonic Mixtures"}

\author{Yi-Ming Duan$^{1}$}
\thanks{These authors contribute equally to this work.}

\author{Liang-Jun He$^{1}$}
\thanks{These authors contribute equally to this work.}

\author{Fabian Maucher$^{2}$}

\author{Yong-Chang Zhang$^{1}$}
\email{zhangyc@xjtu.edu.cn}

\affiliation{$^1$MOE Key Laboratory for Nonequilibrium Synthesis and Modulation of Condensed Matter, Shaanxi Key Laboratory of Quantum Information and Quantum Optoelectronic Devices, School of Physics, Xi'an Jiaotong University, Xi'an 710049, People's Republic of China\\
$^2$Faculty of Mechanical Engineering; Department of Precision and Microsystems Engineering, Delft University of Technology, 2628 CD, Delft, The Netherlands}

\maketitle

\onecolumngrid

\vspace{-10mm}

\section{I. system model}\label{system model}
In this section, we present detailed derivations of effective soft-core interactions in the two-component Rydberg-dressed quantum gases as depicted in Fig.~1 of the main text. We begin with the following Heisenberg equations that describe the coupled light-atom system (see Fig.~1) under the rotating wave approximation, 
\begin{subequations}
	\begin{align}
		i\hbar\frac{\partial}{\partial t}\psi_{g1}&=-\frac{\hbar^{2}}{2M}\nabla^2\psi_{g1}-\hbar\Omega_{p1}\psi_{e1} \label{eq:s1a}\\
		i\hbar\frac{\partial}{\partial t}\psi_{g2}&=-\frac{\hbar^{2}}{2M}\nabla^2\psi_{g2}-\hbar\Omega_{p2}\psi_{e2} \label{eq:s1b}\\
		i\hbar\frac{\partial}{\partial t}\psi_{e1}&=-\frac{\hbar^{2}}{2M}\nabla^2\psi_{e1}-\hbar \Delta_{p1}\psi_{e1}-\hbar\Omega_{p1}\psi_{g1}-\hbar\Omega_{c1}\psi_{r1} \label{eq:s1c} \\
		i\hbar\frac{\partial}{\partial t}\psi_{e2}&=-\frac{\hbar^{2}}{2M}\nabla^2\psi_{e2}-\hbar \Delta_{p2}\psi_{e2}-\hbar\Omega_{p2}\psi_{g2}-\hbar\Omega_{c2}\psi_{r2} \label{eq:s1d}
	\end{align}
	\vspace{-5mm}
	\begin{equation}
		\begin{split}
			i\hbar\frac{\partial}{\partial t}\psi_{r1}=&-\frac{\hbar^{2}}{2M}\nabla^2\psi_{r1}-\hbar (\Delta_{c1}+\Delta_{p1})\psi_{r1}-\hbar\Omega_{c1}\psi_{e1}\\
			&+\int d^3 {\mathbf{r^{\prime}}}U_{11}({\mathbf{r,r^{\prime}}})\psi_{r1}^{\dagger}(\mathbf{r^{\prime}})\psi_{r1}(\mathbf{r^{\prime}})\psi_{r1}
			+\int d^3 {\mathbf{r^{\prime}}}U_{12}({\mathbf{r,r^{\prime}}})\psi_{r2}^{\dagger}(\mathbf{r^{\prime}})\psi_{r2}(\mathbf{r^{\prime}})\psi_{r1}
		\end{split}
		\label{eq:s1e}
	\end{equation}
	\vspace{-5mm}
	\begin{equation}
		\begin{split}
			i\hbar\frac{\partial}{\partial t}\psi_{r2}=&-\frac{\hbar^{2}}{2M}\nabla^2\psi_{r2}-\hbar (\Delta_{c2}+\Delta_{p2})\psi_{r2}-\hbar\Omega_{c2}\psi_{e2}\\
			&+\int d^3 {\mathbf{r^{\prime}}}U_{22}({\mathbf{r,r^{\prime}}})\psi_{r2}^{\dagger}(\mathbf{r^{\prime}})\psi_{r2}(\mathbf{r^{\prime}})\psi_{r2}
			+\int d^3 {\mathbf{r^{\prime}}}U_{21}({\mathbf{r,r^{\prime}}})\psi_{r1}^{\dagger}(\mathbf{r^{\prime}})\psi_{r1}(\mathbf{r^{\prime}})\psi_{r2}
		\end{split}
		\label{eq:s1f}
	\end{equation}
	\label{eq:s1}%
\end{subequations}
where $M$ is the atomic mass, $U_{\alpha\beta}({\mathbf{r,r^{\prime}}})=\frac{\hbar C^{(\alpha\beta)}_{6}}{(\mathbf{r-r^{\prime}})^6}$ ($\alpha,\beta=1,2$) represent the van der Waals interactions between the Rydberg excited states, and $\{ \psi_{g\alpha}, \psi_{e\alpha}, \psi_{r\alpha}\}$ correspond to the field operators of the ground-state, excited-state, and Rydberg-state atoms, respectively. Moreover, $\{ \Delta_{c\alpha}, \Delta_{p\alpha} \}$ and $\{ \Omega_{c\alpha}, \Omega_{p\alpha} \}$ are the detunings between lights and atoms and the Rabi frequencies of the driving fields as illustrated in Fig.~1 in the main text.

In the case of large detunings, it is reasonable to neglect the kinetic energy terms in Eqs.~\eqref{eq:s1c} and~\eqref{eq:s1d} and adiabatically eliminate the excited-state dynamics, yielding
\begin{subequations}
	\begin{align}
		\psi_{e1}=&-\frac{\Omega_{p1}}{\Delta_{p1}}\psi_{g1}-\frac{\Omega_{c1}}{\Delta_{p1}}\psi_{r1} \label{eq:s2a}\\
		\psi_{e2}=&-\frac{\Omega_{p2}}{\Delta_{p2}}\psi_{g2}-\frac{\Omega_{c2}}{\Delta_{p2}}\psi_{r2} \label{eq:s2b}
	\end{align}
	\label{eq:s2}%
\end{subequations}
Substituting the above adiabatic solutions for $\psi_{e1}$ and $\psi_{e2}$ into Eqs.~\eqref{eq:s1a},~\eqref{eq:s1b},~\eqref{eq:s1e}, and~\eqref{eq:s1f} gives
\begin{subequations}
	\begin{align}
		i\hbar\frac{\partial}{\partial t}\psi_{g1}=&-\frac{\hbar^{2}}{2M}\nabla^2\psi_{g1}+\hbar\frac{\Omega^2_{p1}}{\Delta_{p1}}\psi_{g1}+\frac{\hbar\Omega_{p1}\Omega_{c1}}{\Delta_{p1}}\psi_{r1} \label{eq:s3a} \\
		i\hbar\frac{\partial}{\partial t}\psi_{g2}=&-\frac{\hbar^{2}}{2M}\nabla^2\psi_{g2}+\hbar\frac{\Omega^2_{p2}}{\Delta_{p2}}\psi_{g2}+\frac{\hbar\Omega_{p2}\Omega_{c2}}{\Delta_{p2}}\psi_{r2} \label{eq:s3b}
	\end{align}
	\begin{equation}
		\begin{split}
			i\hbar\frac{\partial}{\partial t}\psi_{r1}=&-\frac{\hbar^{2}}{2M}\nabla^2\psi_{r1}+\hbar\frac{\Omega^2_{c1}-\Delta_{p1}(\Delta_{c1}+\Delta_{p1})}{\Delta_{p1}}\psi_{r1}+\hbar\frac{\Omega_{p1}\Omega_{c1}}{\Delta_{p1}}\psi_{g1}\\
			&+\int d^3 {\mathbf{r^{\prime}}}U_{11}({\mathbf{r,r^{\prime}}})\psi_{r1}^{\dagger}(\mathbf{r^{\prime}})\psi_{r1}(\mathbf{r^{\prime}})\psi_{r1}
			+\int d^3 {\mathbf{r^{\prime}}}U_{12}({\mathbf{r,r^{\prime}}})\psi_{r2}^{\dagger}(\mathbf{r^{\prime}})\psi_{r2}(\mathbf{r^{\prime}})\psi_{r1}
		\end{split}
		\label{eq:s3c}
	\end{equation}
	\begin{equation}
		\begin{split}
			i\hbar\frac{\partial}{\partial t}\psi_{r2}=&-\frac{\hbar^{2}}{2M}\nabla^2\psi_{r2}+\hbar\frac{\Omega^2_{c2}-\Delta_{p2}(\Delta_{c2}+\Delta_{p2})}{\Delta_{p2}}\psi_{r2}+\hbar\frac{\Omega_{p2}\Omega_{c2}}{\Delta_{p2}}\psi_{g2}\\
			&+\int d^3 {\mathbf{r^{\prime}}}U_{22}({\mathbf{r,r^{\prime}}})\psi_{r2}^{\dagger}(\mathbf{r^{\prime}})\psi_{r2}(\mathbf{r^{\prime}})\psi_{r2}
			+\int d^3 {\mathbf{r^{\prime}}}U_{21}({\mathbf{r,r^{\prime}}})\psi_{r1}^{\dagger}(\mathbf{r^{\prime}})\psi_{r1}(\mathbf{r^{\prime}})\psi_{r2}
		\end{split}
		\label{eq:s3d}
	\end{equation}
	\label{eq:s3}%
\end{subequations}
where the intermediate excited states have been adiabatically eliminated from the coupled equations. 

Similarly, for sufficiently large effective detunings $\frac{\Omega^2_{c1}-\Delta_{p1}(\Delta_{c1}+\Delta_{p1})}{\Delta_{p1}}$ and $\frac{\Omega^2_{c2}-\Delta_{p2}(\Delta_{c2}+\Delta_{p2})}{\Delta_{p2}}$, we can again neglect the kinetic terms in Eqs.~\eqref{eq:s3c} and~\eqref{eq:s3d} and adiabatically eliminate the Rydberg-state dynamics, which gives
\begin{subequations}
	\begin{equation}
		\begin{split}
			\psi_{r1}=&\frac{\Omega_{p1}\Omega_{c1}}{\Delta_{p1}(\Delta_{c1}+\Delta_{p1})-\Omega^2_{c1}}\psi_{g1}
			+\frac{\Delta_{p1}}{\hbar(\Delta_{p1}(\Delta_{c1}+\Delta_{p1})-\Omega^2_{c1})}
			\int d^3 {\mathbf{r^{\prime}}}U_{11}({\mathbf{r,r^{\prime}}})\psi_{r1}^{\dagger}(\mathbf{r^{\prime}})\psi_{r1}(\mathbf{r^{\prime}})\psi_{r1}\\
			&+\frac{\Delta_{p1}}{\hbar(\Delta_{p1}(\Delta_{c1}+\Delta_{p1})-\Omega^2_{c1})}
			\int d^3 {\mathbf{r^{\prime}}}U_{12}({\mathbf{r,r^{\prime}}})\psi_{r2}^{\dagger}(\mathbf{r^{\prime}})\psi_{r2}(\mathbf{r^{\prime}})\psi_{r1}
		\end{split}
		\label{eq:s4a}
	\end{equation}
	\begin{equation}
		\begin{split}
			\psi_{r2}=&\frac{\Omega_{p2}\Omega_{c2}}{\Delta_{p2}(\Delta_{c2}+\Delta_{p2})-\Omega^2_{c2}}\psi_{g2}
			+\frac{\Delta_{p2}}{\hbar(\Delta_{p2}(\Delta_{c2}+\Delta_{p2})-\Omega^2_{c2})}
			\int d^3 {\mathbf{r^{\prime}}}U_{22}({\mathbf{r,r^{\prime}}})\psi_{r2}^{\dagger}(\mathbf{r^{\prime}})\psi_{r2}(\mathbf{r^{\prime}})\psi_{r2}\\
			&+\frac{\Delta_{p2}}{\hbar(\Delta_{p2}(\Delta_{c2}+\Delta_{p2})-\Omega^2_{c2})}
			\int d^3 {\mathbf{r^{\prime}}}U_{21}({\mathbf{r,r^{\prime}}})\psi_{r1}^{\dagger}(\mathbf{r^{\prime}})\psi_{r1}(\mathbf{r^{\prime}})\psi_{r2}
		\end{split}
		\label{eq:s4b}
	\end{equation}
	\label{eq:s4}
\end{subequations}
By plugging the above adiabatic solutions for $\psi_{r1}$ and $\psi_{r1}$ into Eqs.~\eqref{eq:s3a} and~\eqref{eq:s3b}, one can obtain
\begin{subequations}
	\begin{equation}
		\begin{split}
			i\hbar\frac{\partial}{\partial t}\psi_{g1}=&-\frac{\hbar^{2}}{2M}\nabla^2\psi_{g1}+\hbar\frac{\Omega^2_{p1}}{\Delta_{p1}}\psi_{g1}
			+\frac{\hbar\Omega^2_{p1}\Omega^2_{c1}}{\Delta_{p1}\left( \Delta_{p1}(\Delta_{c1}+\Delta_{p1})-\Omega^2_{c1}\right) }\psi_{g1}\\
			&+\frac{\Omega_{p1}\Omega_{c1}}{(\Delta_{p1}(\Delta_{c1}+\Delta_{p1})-\Omega^2_{c1})}
			\int d^3 {\mathbf{r^{\prime}}}U_{11}({\mathbf{r,r^{\prime}}})\psi_{r1}^{\dagger}(\mathbf{r^{\prime}})\psi_{r1}(\mathbf{r^{\prime}})\psi_{r1}\\
			&+\frac{\Omega_{p1}\Omega_{c1}}{(\Delta_{p1}(\Delta_{c1}+\Delta_{p1})-\Omega^2_{c1})}
			\int d^3 {\mathbf{r^{\prime}}}U_{12}({\mathbf{r,r^{\prime}}})\psi_{r2}^{\dagger}(\mathbf{r^{\prime}})\psi_{r2}(\mathbf{r^{\prime}})\psi_{r1}  
		\end{split}
		\label{eq:s5a}
	\end{equation}
	\begin{equation}
		\begin{split}
			i\hbar\frac{\partial}{\partial t}\psi_{g2}=&-\frac{\hbar^{2}}{2M}\nabla^2\psi_{g2}+\hbar\frac{\Omega^2_{p2}}{\Delta_{p2}}\psi_{g2}
			+\frac{\hbar\Omega^2_{p2}\Omega^2_{c2}}{\Delta_{p2}\left( \Delta_{p2}(\Delta_{c2}+\Delta_{p2})-\Omega^2_{c2}\right) }\psi_{g2}\\
			&+\frac{\Omega_{p2}\Omega_{c2}}{(\Delta_{p2}(\Delta_{c2}+\Delta_{p2})-\Omega^2_{c2})}
			\int d^3 {\mathbf{r^{\prime}}}U_{22}({\mathbf{r,r^{\prime}}})\psi_{r2}^{\dagger}(\mathbf{r^{\prime}})\psi_{r2}(\mathbf{r^{\prime}})\psi_{r2}\\
			&+\frac{\Omega_{p2}\Omega_{c2}}{(\Delta_{p2}(\Delta_{c2}+\Delta_{p2})-\Omega^2_{c2})}
			\int d^3 {\mathbf{r^{\prime}}}U_{21}({\mathbf{r,r^{\prime}}})\psi_{r1}^{\dagger}(\mathbf{r^{\prime}})\psi_{r1}(\mathbf{r^{\prime}})\psi_{r2}  
		\end{split}
		\label{eq:s5b}
	\end{equation}
	\label{eq:s5}%
\end{subequations}
which couples the ground-state operators with the third-order correlalors of the Rydberg-state operators, i.e., $\psi_{r\alpha}^{\dagger}(\mathbf{r^{\prime}})\psi_{r\alpha}(\mathbf{r^{\prime}})\psi_{r\beta}(\mathbf{r})$ ($\alpha,\beta=1,2$). In order to obtain closed equations for the ground-state dynamics, we proceed by considering the dynamics of these third-order correlators, e.g., in Eq.~\eqref{eq:s5a}, as follows,
\begin{subequations}
	\begin{equation}
		i\hbar\frac{\partial}{\partial t}(\psi_{r1}^{\dagger}(\mathbf{r^{\prime}})\psi_{r1}(\mathbf{r^{\prime}})\psi_{r1})=i\hbar\frac{\partial}{\partial t}(\psi_{r1}^{\dagger}(\mathbf{r^{\prime}}))\psi_{r1}(\mathbf{r^{\prime}})\psi_{r1}
		+\psi_{r1}^{\dagger}(\mathbf{r^{\prime}})i\hbar\frac{\partial}{\partial t}(\psi_{r1}(\mathbf{r^{\prime}}))\psi_{r1}+\psi_{r1}^{\dagger}(\mathbf{r^{\prime}})\psi_{r1}(\mathbf{r^{\prime}})i\hbar\frac{\partial}{\partial t}(\psi_{r1})
		\label{eq:s6a}
	\end{equation}
	\begin{equation}
		i\hbar\frac{\partial}{\partial t}(\psi_{r2}^{\dagger}(\mathbf{r^{\prime}})\psi_{r2}(\mathbf{r^{\prime}})\psi_{r1})=i\hbar\frac{\partial}{\partial t}(\psi_{r2}^{\dagger}(\mathbf{r^{\prime}}))\psi_{r2}(\mathbf{r^{\prime}})\psi_{r1}
		+\psi_{r2}^{\dagger}(\mathbf{r^{\prime}})i\hbar\frac{\partial}{\partial t}(\psi_{r2}(\mathbf{r^{\prime}}))\psi_{r1}+\psi_{r2}^{\dagger}(\mathbf{r^{\prime}})\psi_{r2}(\mathbf{r^{\prime}})i\hbar\frac{\partial}{\partial t}(\psi_{r1})
		\label{eq:s6b}
	\end{equation}
	\label{eq:s6}%
\end{subequations}
Substituting Eqs.~\eqref{eq:s3c} and~\eqref{eq:s3d} into the above equations and neglecting the higher-order terms, one can get
\begin{subequations}
	\begin{equation}
		\begin{split}
			&\left[ \hbar\frac{\Omega^2_{c1}-\Delta_{p1}(\Delta_{c1}+\Delta_{p1})}{\Delta_{p1}}+U_{11}({\mathbf{r,r^{\prime}}})\right] \psi_{r1}^{\dagger}(\mathbf{r^{\prime}})\psi_{r1}(\mathbf{r^{\prime}})\psi_{r1}=\\
			&\qquad-\psi_{r1}^{\dagger}(\mathbf{r^{\prime}})\psi_{r1}(\mathbf{r^{\prime}})\int d^3 {\mathbf{r^{\prime\prime}}}U_{12}({\mathbf{r,r^{\prime\prime}}})\psi_{r2}^{\dagger}(\mathbf{r^{\prime\prime}})\psi_{r2}(\mathbf{r^{\prime\prime}})\psi_{r1}\\
			&\qquad+\hbar\frac{\Omega_{p1}\Omega_{c1}}{\Delta_{p1}}\psi_{g1}^{\dagger}(\mathbf{r^{\prime}})\psi_{r1}(\mathbf{r^{\prime}})\psi_{r1}-\hbar\frac{\Omega_{p1}\Omega_{c1}}{\Delta_{p1}}\psi_{r1}^{\dagger}(\mathbf{r^{\prime}})\psi_{g1}(\mathbf{r^{\prime}})\psi_{r1}
			-\hbar\frac{\Omega_{p1}\Omega_{c1}}{\Delta_{p1}}\psi_{r1}^{\dagger}(\mathbf{r^{\prime}})\psi_{r1}(\mathbf{r^{\prime}})\psi_{g1}
		\end{split}
		\label{eq:s7a}
	\end{equation}
	\begin{equation}
		\begin{split}
			&\left[ \hbar\frac{\Omega^2_{c1}-\Delta_{p1}(\Delta_{c1}+\Delta_{p1})}{\Delta_{p1}}+U_{12}({\mathbf{r,r^{\prime}}})\right] \psi_{r2}^{\dagger}(\mathbf{r^{\prime}})\psi_{r2}(\mathbf{r^{\prime}})\psi_{r1}=\\
			&\quad \qquad \hbar\frac{\Omega_{p2}\Omega_{c2}}{\Delta_{p2}}\psi_{g2}^{\dagger}(\mathbf{r^{\prime}})\psi_{r2}(\mathbf{r^{\prime}})\psi_{r1}
			-\hbar\frac{\Omega_{p2}\Omega_{c2}}{\Delta_{p2}}\psi_{r2}^{\dagger}(\mathbf{r^{\prime}})\psi_{g2}(\mathbf{r^{\prime}})\psi_{r1}
			-\hbar\frac{\Omega_{p1}\Omega_{c1}}{\Delta_{p1}}\psi_{r2}^{\dagger}(\mathbf{r^{\prime}})\psi_{r2}(\mathbf{r^{\prime}})\psi_{g1}
		\end{split}
		\label{eq:s7b}
	\end{equation}
	\label{eq:s7}%
\end{subequations}
where we have neglected the kinetic energy term and adiabatically eliminated the correlator dynamics in a similar manner to that for the first-order operators $\psi_{e\alpha}$ and $\psi_{r\alpha}$. Moreover, we have also used the commutation relation $\left[ \psi_{r1}(\mathbf{r^{\prime}}),\ \psi_{r1}^{\dagger}(\mathbf{r^{\prime\prime}})\right]=\delta\left( \mathbf{r}^{\prime}-\mathbf{r}^{\prime\prime}\right)$. One can note that additional second-order correlators of the Rydberg-state operators appear in Eq.~\eqref{eq:s7}, i.e., $\psi_{r1}(\mathbf{r^{\prime}})\psi_{r1}(\mathbf{r})$, $\psi_{r1}^{\dagger}(\mathbf{r^{\prime}})\psi_{r1}(\mathbf{r})$, $\psi_{r1}^{\dagger}(\mathbf{r^{\prime}})\psi_{r1}(\mathbf{r^{\prime}})$, $\psi_{r2}(\mathbf{r^{\prime}})\psi_{r1}(\mathbf{r})$, $\psi_{r2}^{\dagger}(\mathbf{r^{\prime}})\psi_{r1}(\mathbf{r})$, and $\psi_{r2}^{\dagger}(\mathbf{r^{\prime}})\psi_{r2}(\mathbf{r^{\prime}})$, which can be derived via the same method as above for the third-order correlators [cf. Eq.~\eqref{eq:s6}] as follows,
\begin{subequations}
	\begin{align}
		\psi_{r1}(\mathbf{r^{\prime}})\psi_{r1}(\mathbf{r})=&\frac{\frac{2\hbar\Omega^2_{c1}}{\Delta_{p1}(\Omega^2_{c1}-\Delta_{p1}(\Delta_{c1}+\Delta_{p1}))}}{\frac{2\hbar(\Omega^2_{c1}-\Delta_{p1}(\Delta_{c1}+\Delta_{p1}))}{\Delta_{p1}}+U_{11}({\mathbf{r,r^{\prime}}})}\Omega^2_{p1}\psi_{g1}(\mathbf{r^{\prime}})\psi_{g1}(\mathbf{r}) \label{eq:s8a}\\
		\psi_{r1}^{\dagger}(\mathbf{r^{\prime}})\psi_{r1}(\mathbf{r})=&\frac{\Omega^2_{p1}\Omega^2_{c1}}{(\Delta_{p1}(\Delta_{c1}+\Delta_{p1})-\Omega^2_{c1})^2}\psi_{g1}^{\dagger}(\mathbf{r^{\prime}})\psi_{g1}(\mathbf{r}) \label{eq:s8b}\\
		\psi_{r1}^{\dagger}(\mathbf{r^{\prime}})\psi_{r1}(\mathbf{r^{\prime}})=&\frac{\Omega^2_{p1}\Omega^2_{c1}}{(\Delta_{p1}(\Delta_{c1}+\Delta_{p1})-\Omega^2_{c1})^2}\psi_{g1}^{\dagger}(\mathbf{r^{\prime}})\psi_{g1}(\mathbf{r^{\prime}}) \label{eq:s8c} \\
		\psi_{r2}(\mathbf{r^{\prime}})\psi_{r1}(\mathbf{r})=&\frac{\frac{\hbar\Omega_{c1}\Omega_{c2}\Omega_{p1}\Omega_{p2}}{\Delta_{p2}(\Omega^2_{c1}-\Delta_{p1}(\Delta_{c1}+\Delta_{p1}))}+\frac{\hbar\Omega_{c1}\Omega_{c2}\Omega_{p1}\Omega_{p2}}{\Delta_{p1}(\Omega^2_{c2}-\Delta_{p2}(\Delta_{c2}+\Delta_{p2}))}}{\hbar\frac{\Omega^2_{c1}-\Delta_{p1}(\Delta_{c1}+\Delta_{p1})}{\Delta_{p1}}+\hbar\frac{\Omega^2_{c2}-\Delta_{p2}(\Delta_{c2}+\Delta_{p2})}{\Delta_{p2}}+U_{12}({\mathbf{r,r^{\prime}}})} \psi_{g2}(\mathbf{r^{\prime}})\psi_{g1}(\mathbf{r}) \label{eq:s8d} \\
		\psi^{\dagger}_{r2}(\mathbf{r^{\prime}})\psi_{r1}(\mathbf{r})=&\frac{\Omega_{c1}\Omega_{p1}}{\Delta_{p1}(\Delta_{c1}+\Delta_{p1})-\Omega^2_{c1}}\frac{\Omega_{c2}\Omega_{p2}}{\Delta_{p2}(\Delta_{c2}+\Delta_{p2})-\Omega^2_{c2}} \psi^{\dagger}_{g2}(\mathbf{r^{\prime}})\psi_{g1}(\mathbf{r}) \label{eq:s8e}\\
		\psi^{\dagger}_{r2}(\mathbf{r^{\prime}})\psi_{r2}(\mathbf{r^{\prime}})=&\frac{\Omega_{c2}\Omega_{p2}}{\Delta_{p2}(\Delta_{c2}+\Delta_{p2})-\Omega^2_{c2}}\frac{\Omega_{c2}\Omega_{p2}}{\Delta_{p2}(\Delta_{c2}+\Delta_{p2})-\Omega^2_{c2}} \psi^{\dagger}_{g2}(\mathbf{r^{\prime}})\psi_{g2}(\mathbf{r^{\prime}}) \label{eq:s8f}
	\end{align}
	\label{eq:s8}
\end{subequations}

By substituting Eqs.~\eqref{eq:s7} and~\eqref{eq:s8} back into Eq.~\eqref{eq:s5a}, we reach the following equation for $\psi_{g1}$ involving only the ground-state field operators,
\begin{equation}
	\begin{split}
		i\hbar\frac{\partial}{\partial t}\psi_{g1}=&-\frac{\hbar^{2}}{2M}\nabla^2\psi_{g1}
		+\int d^3 {\mathbf{r^{\prime}}}	\frac{\frac{2\hbar\Omega^4_{c1}\Omega^4_{p1}{C}_6^{(11)}}{(\Omega^2_{c1}-\Delta_{p1}(\Delta_{c1}+\Delta_{p1}))^4}}{\frac{\Delta_{p1}{C}_6^{(11)}}{2(\Omega^2_{c1}-\Delta_{p1}(\Delta_{c1}+\Delta_{p1}))}+{(\mathbf{r}-\mathbf{r^\prime})^6}} \psi_{g1}^{\dagger}(\mathbf{r^{\prime}})\psi_{g1}(\mathbf{r^{\prime}})\psi_{g1} \\
		&+\int d^3 {\mathbf{r^{\prime}}}\frac{\frac{\hbar\Omega^2_{c1}\Omega^2_{c2}\Omega^2_{p1}\Omega^2_{p2}{C}_6^{(12)}}{(\Omega^2_{c1}-\Delta_{p1}(\Delta_{c1}+\Delta_{p1}))^2(\Omega^2_{c2}-\Delta_{p2}(\Delta_{c2}+\Delta_{p2}))^2} }{\frac{\Delta_{p1}\Delta_{p2}{C}_6^{(12)}}{ \Delta_{p2}(\Omega^2_{c1}-\Delta_{p1}(\Delta_{c1}+\Delta_{p1})) + \Delta_{p1}(\Omega^2_{c2}-\Delta_{p2}(\Delta_{c2}+\Delta_{p2}))}+ (\mathbf{r}-\mathbf{r^\prime})^6 } \psi_{g2}^{\dagger}(\mathbf{r^{\prime}})\psi_{g2}(\mathbf{r^{\prime}})\psi_{g1}		
	\end{split}
	\label{eq:s9}
\end{equation}
where the constant terms have been omitted. Similarly, through repeating the above derivations for Eq.~\eqref{eq:s5b}, we obtain the other equation for $\psi_{g2}$, i.e.,
\begin{equation}
	\begin{split}
		i\hbar\frac{\partial}{\partial t}\psi_{g2}=&-\frac{\hbar^{2}}{2M}\nabla^2\psi_{g2}
		+\int d^3 {\mathbf{r^{\prime}}}	\frac{\frac{2\hbar\Omega^4_{c2}\Omega^4_{p2}{C}_6^{(22)}}{(\Omega^2_{c2}-\Delta_{p2}(\Delta_{c2}+\Delta_{p2}))^4}}{\frac{\Delta_{p2}{C}_6^{(22)}}{2(\Omega^2_{c2}-\Delta_{p2}(\Delta_{c2}+\Delta_{p2}))}+{(\mathbf{r}-\mathbf{r^\prime})^6}} \psi_{g2}^{\dagger}(\mathbf{r^{\prime}})\psi_{g2}(\mathbf{r^{\prime}})\psi_{g2} \\
		&+\int d^3 {\mathbf{r^{\prime}}}\frac{\frac{\hbar\Omega^2_{c1}\Omega^2_{c2}\Omega^2_{p1}\Omega^2_{p2}{C}_6^{(12)}}{(\Omega^2_{c1}-\Delta_{p1}(\Delta_{c1}+\Delta_{p1}))^2(\Omega^2_{c2}-\Delta_{p2}(\Delta_{c2}+\Delta_{p2}))^2} }{\frac{\Delta_{p1}\Delta_{p2}{C}_6^{(12)}}{ \Delta_{p2}(\Omega^2_{c1}-\Delta_{p1}(\Delta_{c1}+\Delta_{p1})) + \Delta_{p1}(\Omega^2_{c2}-\Delta_{p2}(\Delta_{c2}+\Delta_{p2}))}+ (\mathbf{r}-\mathbf{r^\prime})^6 } \psi_{g1}^{\dagger}(\mathbf{r^{\prime}})\psi_{g1}(\mathbf{r^{\prime}})\psi_{g2}		
	\end{split}
	\label{eq:s10}
\end{equation}

Eventually, we have arrived at the above closed Eqs.~\eqref{eq:s9} and~\eqref{eq:s10}, which describes the dynamics of the atoms in the two hyperfine ground states, subjected to effective long-range interactions that arise from the van der Waals interactions between Rydberg excited states. By employing mean-field theory and replacing the field operators with their expected values $\Psi_\alpha=\langle \psi_{g\alpha}\rangle$, Eqs.~\eqref{eq:s9} and~\eqref{eq:s10} can be rewritten as a set of coupled equations for the mean-field wave functions of each component,
\begin{equation}
	\begin{split}
		i\hbar \frac{\partial}{\partial t}\Psi_\alpha (\mathbf{r})&=-\frac{\hbar^{2} \nabla^{2}}{2 M}{\Psi_\alpha}(\mathbf{r})
		+\sum_{\beta=1,2}\left[ g_{\alpha\beta}\left|\Psi_\beta (\mathbf{r}) \right|^2+\int \mathrm{d}^{3} \mathbf{r}'
		V_{\alpha\beta}\left(\mathbf{r}-\mathbf{r}'\right)\left|\Psi_\beta\left(\mathbf{r}'\right)\right|^2\right]\Psi_\alpha(\mathbf{r})
	\end{split}
	\label{eq:s11}
\end{equation}
i.e., the Gross-Pitaevskii equation in the main text, where we have added the contact interactions and defined
\begin{subequations}
	\begin{align}
		V_{\alpha\beta}(\mathbf{r})=&\frac{\tilde{C}_6^{(\alpha\beta)}}{1+\left(\frac{\mathbf{r}}{R_c^{(\alpha\beta)}}\right)^6}\qquad (\alpha, \beta=1,2)\\
		\tilde{C}_6^{(\alpha\beta)}=&\frac{\hbar\Omega^2_{c\alpha}\Omega^2_{c\beta}\Omega^2_{p\alpha}\Omega^2_{p\beta}  }{\Delta_{p\alpha}\Delta_{p\beta}}
		\frac{  \Delta_{p\beta}(\Omega^2_{c\alpha}-\Delta_{p\alpha}(\Delta_{c\alpha}+\Delta_{p\alpha})) + \Delta_{p\alpha}(\Omega^2_{c\beta}-\Delta_{p\beta}(\Delta_{c\beta}+\Delta_{p\beta})) }{(\Omega^2_{c\alpha}-\Delta_{p\alpha}(\Delta_{c\alpha}+\Delta_{p\alpha}))^2(\Omega^2_{c\beta}-\Delta_{p\beta}(\Delta_{c\beta}+\Delta_{p\beta}))^2}\\
		R_c^{(\alpha\beta)}=&\left( \frac{C_6^{(\alpha\beta)}}{ \frac{\Omega^2_{c\alpha}-\Delta_{p\alpha}(\Delta_{c\alpha}+\Delta_{p\alpha})}{\Delta_{p\alpha}} + \frac{\Omega^2_{c\beta}-\Delta_{p\beta}(\Delta_{c\beta}+\Delta_{p\beta})}{\Delta_{p\beta}}}\right) ^{1/6}
	\end{align}
	\label{eq:s12}
\end{subequations}

To experimentally implement this binary Rydberg-dressed Bosonic mixtures satisfying the constraints $\tilde{C}_6^{(11)}=\tilde{C}_6^{(22)}=\tilde{C}_6$, $R_c^{(11)}=R_c^{(22)}=R_c$, we propose potential configurations as specified in Table~\ref{tab1}, where the van der Waals interaction coefficient $C_6$ is calculated using the ARC-open source library~\cite{Nikola2017ARC,Nikola2021ARC}. Here we present two examples of rubidium and caesium atoms. For the caesium atoms, we choose the $n_1 S_{1/2}$ and $n_2 S_{1/2}$ states with close principal quantum numbers as the two Rydberg excited states. In this configuration, the inter- and intra-component soft-core interactions are all isotropic. For the rubidium atoms, to meet the aforementioned constraints, it is hard to find two appropriate $S$-orbital states, alternatively, one can use, e.g., $n_1 S_{1/2}$ and $n_2 D_{1/2}$ states as the two Rydberg excited states. However, in this case, the $D$-orbital component would experience an anisotropic soft-core interaction, i.e., the effective strength is no longer fixed and varies with spatial direction. Despite of this anisotropy, it becomes approximately isotropic in the long-range regime~\cite{HenkelPhD}.

Although it is difficult to continuously tune $R^{(12)}_c$ across the full range of the phase diagram, we can reach the situations of both $R^{(12)}_c/R_c<1$ and $R^{(12)}_c/R_c>1$ by properly selecting the Rydberg states~\cite{Igor2015NJP,Qian2015PRA,Igor2016NJP,Igor2016PRA,Igor2019PRE} and exquisitely adjusting the Rabi frequencies and detunings (e.g., the relevant parameters provided in Table~\ref{tab1}), permitting the exploration of intriguing symmetry-breaking phenomena in the miscible regime (e.g., segregated tubular/planar state, and other crystal states) as well as in the phase-separation regime. This flexibility in tuning the blockade radii allows for a detailed investigation of  rich phases as well as dynamical behaviours in these two distinct regimes.

\begin{table}
	\setlength{\tabcolsep}{8pt}
	\caption{\textbf{Implementation Schemes}}
	\centering\scalebox{1}{
		\begin{tabular}{ccccc}
			\toprule[0.5mm]
			\thead{Configurations\\ ($n_\alpha$: principal quantum number) \\ ($mj_\alpha$: magnetic quantum number)} & $C_{6}\ \left( \rm{GHz \times \mu m^6}\right)$ & \thead{Adjustable \\Parameters}  $\left( \rm{MHz}\right)$  &  \thead{Dimensionless \\Parameters}  & \thead{Usage in \\the main text}\\ 
			\midrule[0.5mm]
			\thead{atom: Rb $\leftrightarrow$ Rb\\ orbit: S $\leftrightarrow$ D\\$n_{1}=62$; $n_{2}=43$\\$mj_{1}=0.5$; $mj_{2}=0.5$ } & \thead{$C^{(11)}_{6}$=203.89\\$C^{(22)}_{6}$=203.73\\$C^{(12)}_{6}$=0.03} &\thead{$\Omega_{c1}=\Omega_{c2}=2\pi \times 4$ \\ $\Omega_{p1}=\Omega_{p2}=2\pi \times2.4$ \\ $\Delta_{c1}=\Delta_{c2}=-2\pi \times20$ \\ $\Delta_{p1}=\Delta_{p2}=-2\pi \times20$} & \thead{$R^{(12)}_{c}/R_{c}=0.23$\\ 
				$\tilde{C}^{(12)}_{6}/\tilde{C}_{6}=1$
			} 			& ~\\ 
			
			\midrule[0.1mm]
			
			\thead{atom: Rb $\leftrightarrow$ Rb\\ orbit: S $\leftrightarrow$ D\\$n_{1}=48$; $n_{2}=42$\\$mj_{1}=0.5$; $mj_{2}=0.5$ } & \thead{$C^{(11)}_{6}$=9.31\\$C^{(22)}_{6}$=10.21\\$C^{(12)}_{6}$=2.28} &\thead{$\Omega_{c1}=\Omega_{c2}=2\pi \times 4$ \\ $\Omega_{p1}=\Omega_{p2}=2\pi \times4.48$ \\ $\Delta_{c1}=\Delta_{c2}=-2\pi \times20$ \\ $\Delta_{p1}=\Delta_{p2}=-2\pi \times20$} & \thead{$R^{(12)}_{c}/R_{c}=0.78$\\ 
			$\tilde{C}^{(12)}_{6}/\tilde{C}_{6}=1$
			} & ~ \\
			
			\midrule[0.1mm]
			
			\thead{atom: Cs $\leftrightarrow$ Cs\\ orbit: S $\leftrightarrow$ S\\$n_{1}=60$; $n_{2}=62$\\$mj_{1}=0.5$; $mj_{2}=0.5$ } & \thead{$C^{(11)}_{6}$=100.95\\$C^{(22)}_{6}$=150.32\\$C^{(12)}_{6}$=94.12} &\thead{$\Omega_{c1}=\Omega_{c2}=2\pi \times4$ \\ $\Omega_{p1}=\Omega_{p2}=2\pi \times2.48$ \\ $\Delta_{c1}=-2\pi \times20$ \\ $\Delta_{c2}=-2\pi \times44.6$\\ $\Delta_{p1}=-2\pi \times20 $\\$\Delta_{p2}=-2\pi \times14.8 $} & \thead{$R^{(12)}_{c}/R_{c}=0.95$\\ 
				$\tilde{C}^{(12)}_{6}/\tilde{C}_{6}=1.02$} & Fig.~4 \\ 	 
			
			
			\midrule[0.1mm] 
			
			\thead{atom: Cs $\leftrightarrow$ Cs\\ orbit: S $\leftrightarrow$ S\\$n_{1}=68$; $n_{2}=70$\\$mj_{1}=0.5$; $mj_{2}=0.5$ } & \thead{$C^{(11)}_{6}$=458.67\\$C^{(22)}_{6}$=649.75\\$C^{(12)}_{6}$=184.69} &\thead{$\Omega_{c1}=\Omega_{c2}=2\pi \times4$ \\ $\Omega_{p1}=\Omega_{p2}=2\pi \times1.81$ \\ $\Delta_{c1}=-2\pi \times20$ \\ $\Delta_{c2}=-2\pi \times41.2$\\ $\Delta_{p1}=-2\pi \times20 $\\$\Delta_{p2}=-2\pi \times15.4 $} & \thead{$R^{(12)}_{c}/R_{c}=0.83$\\ 
				$\tilde{C}^{(12)}_{6}/\tilde{C}_{6}=1.015$}  & \thead{Case I \\in Fig.~1}\\ 
			
			\midrule[0.1mm]
			
			\thead{atom: Cs $\leftrightarrow$ Cs\\ orbit: S $\leftrightarrow$ S\\$n_{1}=74$; $n_{2}=76$\\$mj_{1}=0.5$; $mj_{2}=0.5$ } & \thead{$C^{(11)}_{6}$=1263.9\\$C^{(22)}_{6}$=1737.6\\$C^{(12)}_{6}$=159} &\thead{$\Omega_{c1}=\Omega_{c2}=2\pi \times4$ \\ $\Omega_{p1}=\Omega_{p2}=2\pi \times1.47$ \\ $\Delta_{c1}=-2\pi \times20$ \\ $\Delta_{c2}=-2\pi \times39.2$\\ $\Delta_{p1}=-2\pi \times20 $\\$\Delta_{p2}=-2\pi \times15.7 $} & \thead{$R^{(12)}_{c}/R_{c}=0.69$\\ 
				$\tilde{C}^{(12)}_{6}/\tilde{C}_{6}=1.013$} & ~ \\
			
			\midrule[0.1mm]
			
			\thead{atom: Cs $\leftrightarrow$ Cs\\ orbit: S $\leftrightarrow$ S\\$n_{1}=75$; $n_{2}=77$\\$mj_{1}=0.5$; $mj_{2}=0.5$ } & \thead{$C^{(11)}_{6}$=1483.6\\$C^{(22)}_{6}$=2035.5\\$C^{(12)}_{6}$=121.9} &\thead{$\Omega_{c1}=\Omega_{c2}=2\pi \times4$ \\ $\Omega_{p1}=\Omega_{p2}=2\pi \times1.42$ \\ $\Delta_{c1}=-2\pi \times20$ \\ $\Delta_{c2}=-2\pi \times38.9$\\ $\Delta_{p1}=-2\pi \times20 $\\$\Delta_{p2}=-2\pi \times15.8 $} & \thead{$R^{(12)}_{c}/R_{c}=0.64$\\ 
				$\tilde{C}^{(12)}_{6}/\tilde{C}_{6}=1.012$} & ~ \\
			
			\midrule[0.1mm]
			
			\thead{atom: Cs $\leftrightarrow$ Cs\\ orbit: S $\leftrightarrow$ S\\$n_{1}=60$; $n_{2}=61$\\$mj_{1}=0.5$; $mj_{2}=0.5$ } & \thead{$C^{(11)}_{6}$=100.95\\$C^{(22)}_{6}$=123.41\\$C^{(12)}_{6}$=252.56} &\thead{$\Omega_{c1}=\Omega_{c2}=2\pi \times4$ \\ $\Omega_{p1}=\Omega_{p2}=2\pi \times2.48$ \\ $\Delta_{c1}=-2\pi \times20$ \\ $\Delta_{c2}=-2\pi \times31.6$ \\ $\Delta_{p1}=-2\pi \times20 $\\$\Delta_{p2}=-2\pi \times17.2 $} & \thead{$R^{(12)}_{c}/R_{c}=1.15$\\ 
				$\tilde{C}^{(12)}_{6}/\tilde{C}_{6}=1.005$} & \thead{Case II \\in Fig.~1} \\			 	 
			\bottomrule[0.5mm]
	\end{tabular}}
	\label{tab1}
\end{table}

\section{II. Variational approach}\label{variational_approach}

In this section, we elaborate in detail on the variational approach utilized in the main text. Although this method provides a relatively rough estimation of the system’s properties when compared with numerical simulations, it offers a straightforward way to predict potential new phases and interpret the numerical results.

Inspired by the numerical results, we approximate the density profile in each lattice site of the crystal states using a Gaussian distribution. In this approach, the width of the Gaussian function and the separations between nearest neighbours serve as variational parameters. By minimizing the total energy for a fixed average density under periodic boundary conditions (i.e., in the thermodynamic limit)~\cite{santi2018freezing,santi2020a,Prestipino_2019}, we can obtain the ground-state properties of the system. This variational method is applicable to a wide range of ground-state profiles discussed in the main text, including the simple cubic (SC), face-centered cubic (FCC), body-centered cubic (BCC) crystals, as well as the segregated tubular (ST) and segregated planar (SP) states.

For simplicity, let us start with a 3D single-component Rydberg-dressed quantum gas, resembling the special situation of $R^{(12)}_c/R_c=1$ considered in the main text, the result of which can be straightforwardly extended to the scenario of binary mixtures. The total energy of this single-component system reads,
\begin{equation}
	E\left[\rho\right]=\int\mathrm{d}^3\mathbf{r}\left[-\frac{1}{2}\sqrt{\rho\left(\mathbf{r}\right)}\nabla^2\sqrt{\rho\left(\mathbf{r}\right)}+\frac{1}{2}g\rho\left(\mathbf{r}\right)^2\right]+\frac{1}{2}\iint\mathrm{d}^3\mathbf{r}\mathrm{d}^3\mathbf{r}'
	V\left(\mathbf{r}-\mathbf{r}'\right)	\rho\left(\mathbf{r}\right)\rho\left(\mathbf{r}'\right),
	\label{eq:s13}
\end{equation}
where $V(\mathbf{r})=\frac{M R^2_c}{\hbar^2}\frac{\tilde{C}_6}{1+\mathbf{r}^6}$, and we have rescaled the energy and space with $\hbar^2/(M R^2_c)$ and $R_c$, respectively. Here $\rho\left(\mathbf{r}\right)$ represents the density distribution. For the BCC state, it can be written as 
\begin{equation}
	\rho_{\mathrm{BCC}}=\frac{a^3\rho_{3\mathrm{D}}}{2\pi^{\frac{3}{2}}\sigma^3}\sum_{l,m,n \in {\rm Integers}}\left[e^{-\frac{\left(x+l a\right)^2+\left(y+m a\right)^2+\left(z+n a\right)^2}{\sigma^2}}+e^{-\frac{\left(x-\frac{1}{2}a+l a\right)^2+\left(y-\frac{1}{2}a+m a\right)^2+\left(z-\frac{1}{2}a+n a\right)^2}{\sigma^2}}\right],
	\label{eq:s14}
\end{equation}
with $\sigma$ being the width of each Gaussian profile, $\rho_{3\mathrm{D}}$ the average density, $a$ the lattice constant. 

\begin{figure}[!b]
	\renewcommand\thefigure{S1}
	\centering
	\includegraphics[width=0.7\columnwidth]{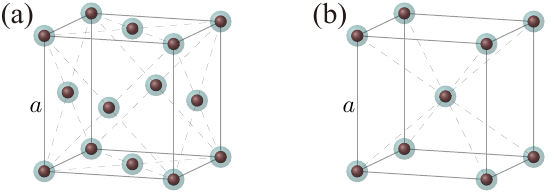}
	\caption{Schematic plots of the spatial density profiles of (a) face-centered cubic (FCC) and (b) body-centered cubic (BCC) states for a single-component system. }
	\label{FCC_BCC}
\end{figure}

Although the kinetic energy is difficult to formulate analytically, it can be calculated numerically. Here, we mainly focus on the interaction energy terms. The contact interaction energy per unit volume is given by
\begin{equation}
	\begin{split}
		\mathcal{E}_{\mathrm{loc}}=\frac{1}{2a^3}\int_{\mathrm{C}}\mathrm{d}^3\mathbf{r}g\rho_{\mathrm{BCC}}^2
		=\frac{a^3 g\rho_{3\mathrm{D}}^2}{4\left(2\pi\right)^{\frac{3}{2}}\sigma^3}\left[\vartheta_3^3\left(e^{-\frac{a^2}{2\sigma^2}}\right)+\vartheta_2^3\left(e^{-\frac{a^2}{2\sigma^2}}\right)\right],
	\end{split}
	\label{eq:s15}
\end{equation}
where $\int_{\mathrm{C}}$ denotes the integral over a single cubic region [cf. Fig.~\ref{FCC_BCC}(b)] and $\vartheta_k$ ($k=1,2,3,4$) is the Jacobi theta function. In order to calculate the soft-core interaction energy, we first derive the density in $\mathbf{k}$-space as follows, 
\begin{equation}
	\begin{split}
		\tilde{\rho}_{\mathrm{BCC}}\left(\mathbf{k}\right)=&\frac{1}{\left(2\pi\right)^{3}}\int\mathrm{d}^3\mathbf{r}e^{-i\mathbf{k}\cdot\mathbf{r}}\rho_{\mathrm{BCC}}\\
		=&\frac{1}{2}a^3\rho_{3\mathrm{D}}e^{-\frac{1}{4}\sigma^2 k^2}
		\lim_{\mathbb{M}\to +\infty}\left[\mathcal{D}_{\mathbb{M}}\left(ak_x\right)\mathcal{D}_{\mathbb{M}}\left(ak_y\right)\mathcal{D}_{\mathbb{M}}\left(ak_z\right)
		+\mathcal{D}_{\mathbb{M}+\frac{1}{2}}\left(ak_x\right)\mathcal{D}_{\mathbb{M}+\frac{1}{2}}\left(ak_y\right)\mathcal{D}_{\mathbb{M}+\frac{1}{2}}\left(ak_z\right)\right]\\
		=&\rho_{3\mathrm{D}}e^{-\frac{1}{4}\sigma^2 k^2} \tilde{\rho}'_{\mathrm{BCC}}\left(\mathbf{k}\right),
	\end{split}
	\label{eq:s16}
\end{equation}
where $\mathcal{D}_{\mathbb{M}}\left(x\right)=\frac{1}{2\pi}\frac{\sin\left(\mathbb{M}+\frac{1}{2}\right)x}{\sin\frac{x}{2}}$ is the Dirichlet kernel and
\begin{equation}
	\begin{split}
		\tilde{\rho}'_{\mathrm{BCC}}\left(\mathbf{k}\right)
		=&\sum_{l,m,n\in {\rm Integers}}\left[\delta\left(k_x+l \frac{4\pi}{a}\right)\delta\left(k_y+m \frac{4\pi}{a}\right) \delta\left(k_z+n \frac{4\pi}{a}\right)\right.\\
		&\left.+\delta\left(k_x-\frac{2\pi}{a}+l \frac{4\pi}{a}\right)\delta\left(k_y-\frac{2\pi}{a}+m \frac{4\pi}{a}\right) \delta\left(k_z+n \frac{4\pi}{a}\right)\right.\\
		&\left.+\delta\left(k_x+l \frac{4\pi}{a}\right)\delta\left(k_y-\frac{2\pi}{a}+m \frac{4\pi}{a}\right) \delta\left(k_z-\frac{2\pi}{a}+n \frac{4\pi}{a}\right)\right.\\
		&\left.+\delta\left(k_x-\frac{2\pi}{a}+l \frac{4\pi}{a}\right)\delta\left(k_y+m \frac{4\pi}{a}\right) \delta\left(k_z-\frac{2\pi}{a}+n \frac{4\pi}{a}\right)\right].
	\end{split}
	\label{eq:s17}
\end{equation}
It is worthy noting that the BCC lattice exhibits an FCC structure in $\mathbf{k}$-space. Subsequently, we calculate the soft-core interaction energy per unit volume by employing the convolution theorem as below,
\begin{equation}
	\begin{split}
		\mathcal{E}_{\mathrm{sof}}=&\frac{1}{2a^3}\int_{\mathrm{C}}\mathrm{d}^3 \mathbf{r}\int\mathrm{d}^3 \mathbf{k}\rho_{\mathrm{BCC}}\tilde{V}\left(\mathbf{k}\right)\tilde{\rho}_{\mathrm{BCC}}\left(\mathbf{k}\right)e^{i\mathbf{k}\cdot\mathbf{r}}\\
		=&\frac{\sqrt{\pi}\rho_{3\mathrm{D}}^2 C_6}{3\sigma^3}\int\mathrm{d}^3 \mathbf{r}\int\mathrm{d}^3 \mathbf{k}e^{\frac{1}{4}\sigma^2 k^2}f\left(k\right) e^{i\mathbf{k}\cdot\mathbf{r}}e^{-\frac{x^2+y^2+z^2}{\sigma^2}}\tilde{\rho}'_{\mathrm{BCC}}\left(\mathbf{k}\right)\\
		=&\frac{\pi^2\rho_{3\mathrm{D}}^2 C_6}{3} \left[1+6\sum_{m=1}^{+\infty}f\left(m\frac{4\pi}{a}\right) +12\sum_{\substack{m_1,m_2=1}}^{+\infty}f\left(\sqrt{m_1^2+m_2^2}\frac{2\pi}{a}\right)\right.\\
		&\left.+8\sum_{m_1,m_2,m_3=1}^{+\infty}f\left(\sqrt{m_1^2+m_2^2+m_3^2}\frac{4\pi}{a}\right) +24\sum_{\substack{m_1,m_2,m_3=1}}^{+\infty}f\left(\sqrt{m_1^2+m_2^2+4m_3^2}\frac{2\pi}{a}\right)\right],
	\end{split}
	\label{eq:s18}
\end{equation}
where $f\left(x\right) \equiv \frac{1}{x} e^{-\frac{x^2+x}{2}} \left[e^{-\frac{x}{2}}-2\cos\left(\frac{\sqrt{3}}{2}x+\frac{\pi}{3}\right)\right]$. Now we have obtained the variational energy of the BCC state. By minimizing it with respect to $a$ and $\sigma$, one can readily determine the lowest energy as well as the corresponding optimal parameters. The calculation for the FCC state follows a similar approach. It just needs to repeat the above calculations by replacing the density profile with
\begin{equation}
	\begin{split}
		\rho_{\mathrm{FCC}}=\frac{a^3\rho_{3\mathrm{D}}}{2\pi^{\frac{3}{2}}\sigma^3}\sum_{l,m,n \in {\rm Integers}}&\left[e^{-\frac{\left(x+l a\right)^2+\left(y+m a\right)^2+\left(z+n a\right)^2}{\sigma^2}}+e^{-\frac{\left(x-\frac{1}{2}a+l a\right)^2+\left(y-\frac{1}{2}a+m a\right)^2+\left(z+n a\right)^2}{\sigma^2}}\right.\\
		&\left. +e^{-\frac{\left(x-\frac{1}{2}a+l a\right)^2+\left(y+m a\right)^2+\left(z-\frac{1}{2}a+n a\right)^2}{\sigma^2}} +e^{-\frac{\left(x+l a\right)^2+\left(y-\frac{1}{2}a+m a\right)^2+\left(z-\frac{1}{2}a+n a\right)^2}{\sigma^2}}\right].
	\end{split}
	\label{eq:s19}
\end{equation}

\begin{figure}[!t]
	\renewcommand\thefigure{S2}
	\centering
	\includegraphics[width=0.8\columnwidth]{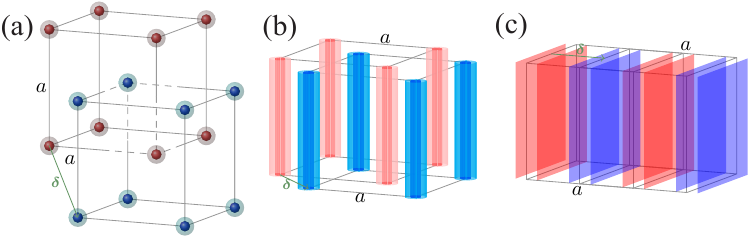}
	\caption{(a) simple cubic (SC), (b) segregated tubular (ST), and (c) segregated planar (SP) states for the two-component Rydberg-dressed quantum gases. Red and blue represent component 1 and component 2, respectively.} 
	\label{figs2}
\end{figure}

Now let us turn to the two-component case, where exotic states including SC, ST, and SP states (see Fig.~\ref{figs2}) emerge in addition to the FCC and BCC crystals. To variationally reproduce the surprising ST and SP profiles, instead of an equal width $\sigma$ for the single-component system discussed above, we assume independent widths of the Gaussian function along different directions. For instance, the density profile of the SC state as shown in Fig.~\ref{figs2}(a) can be approximated as follows,
\begin{subequations}
	\begin{align}
		\rho_{1\mathrm{SC}}=&\frac{a^3\rho_{3\mathrm{D}}}{2\pi^{\frac{3}{2}}\sigma_x\sigma_y\sigma_z} \sum_{l,m,n\in {\rm Integers}}e^{-\left[\frac{\left(x+l a\right)^2}{\sigma_x^2}+\frac{\left(y+m a\right)^2}{\sigma_y^2}+\frac{\left(z+n a\right)^2}{\sigma_z^2}\right]} \\
		\rho_{2\mathrm{SC}}=&\frac{a^3\rho_{3\mathrm{D}}}{2\pi^{\frac{3}{2}}\sigma_x\sigma_y\sigma_z} \sum_{l,m,n\in {\rm Integers}}e^{-\left[\frac{\left(x+l a+\delta_x\right)^2}{\sigma_x^2}+\frac{\left(y+m a+\delta_y\right)^2}{\sigma_y^2}+\frac{\left(z+n a+\delta_z\right)^2}{\sigma_z^2}\right]}
	\end{align}
	\label{eq:s20}%
\end{subequations}
with $(\delta_x,\delta_y,\delta_z)$ being the misalignment vector between the two components. Subsequently, we can calculate the corresponding energy by plugging the above densities into the following energy functional
\begin{equation}
	\begin{split}
		E\left[\rho_1,\rho_2\right]=&\int\mathrm{d}^3\mathbf{r}\left[-\frac{1}{2}\sqrt{\rho_1}\nabla^2\sqrt{\rho_1}-\frac{1}{2}\sqrt{\rho_2}\nabla^2\sqrt{\rho_2}+\frac{1}{2}\left( g_{11}\rho^2_1+g_{22}\rho^2_2+2g_{12}\rho_1\rho_2 \right)\right]\\
		&+\frac{1}{2}\iint\mathrm{d}^3\mathbf{r}\mathrm{d}^3\mathbf{r}' \left[V_{11}\left(\mathbf{r}-\mathbf{r}'\right) \rho_1\left(\mathbf{r}\right)\rho_1\left(\mathbf{r}'\right) +V_{22}\left(\mathbf{r}-\mathbf{r}'\right) \rho_2\left(\mathbf{r}\right)\rho_2\left(\mathbf{r}'\right) +2V_{12}\left(\mathbf{r}-\mathbf{r}'\right) \rho_1\left(\mathbf{r}\right)\rho_2\left(\mathbf{r}'\right) \right].
	\end{split}
	\label{eq:s21}%
\end{equation}
The remaining calculations follow the same procedure as in the single-component case, and will therefore not be repeated here. By minimizing the final energy with respect to $\sigma_x$, $\sigma_y$, $\sigma_z$, and $a$, one can obtain stable states with minimum energies at optimal parameters. When $\sigma_x$, $\sigma_y$, and $\sigma_z$ are all finite and equal at the optimal points, it corresponds to the profile of the SC states. However, if one (two) of them becomes divergent, it recovers the unexpected distribution of the segregated tubular (planar) state as exhibited in Fig.~\ref{figs2}.

\section{III. Density correlation function}
In this section, we detail the calculation of the density correlation function displayed in Fig.~4 in the main text. The density correlation function is defined as follows~\cite{PhysRevLett.95.248301}:
\begin{equation}
	g(r)=\braket{\rho(\mathbf{r}_1)\rho(\mathbf{r}_2)}/\rho_{\mathrm{3D}}^2,
\end{equation}
where $\braket{\cdot}$ denotes the average of the product $\rho(\mathbf{r}_1)\rho(\mathbf{r}_2)$ over all positions $\mathbf{r}_1$ and $\mathbf{r}_2$ in 3D space that satisfy $|\mathbf{r}_1-\mathbf{r}_2|=r$, and $\rho_{\mathrm{3D}}$ represents the 3D average density of the condensate. In the case of an unmodulated state, the density is a position-independent constant, i.e., $\rho(\mathbf{r})=\rho_{\rm 3D}$, and thus $g(r)=1$. However, for a modulated state that presents lattice structures, $g(r)$ oscillates with $r$ and eventually converges to $1$ at large separation, where the oscillation periodicity reflects the lattice constant. Therefore, the density correlation function serves as an indicator describing the process of establishing modulated density distributions during the dynamics of the quantum gas. 

For the segregated tubular state, spontaneous spatial symmetry-breaking occurs only in the two-dimensional plane perpendicular to the tube direction. That is, the density along the tube direction is homogeneous. Hence, to describe the self-organization process, it needs only to consider the correlation function $g_\perp(r)$ in the transverse plane as below,
\begin{equation}
	g_{\perp}(r)=\braket{\rho_{\perp}(\mathbf{r}_1)\rho_{\perp}(\mathbf{r}_2)}/\rho_{\mathrm{3D}}^2.
\end{equation}
Here, we have assumed that the tubes are parallel to the $z$ axis, $\rho_{\perp}(\mathbf{r})=L^{-1}_z\int^{L_z}_0\rho(x,y,z)\mathrm{d}z$ is the average density profile in the $x\text{-}y$ plane with $L_z$ being the length of the numerical box along $z$ direction and $\mathbf{r}_i\ (i=1,2)$ the position coordinates in the transverse plane. Although the tube direction is actually random in realistic simulations, it can be either artificially rotated to the $z$ direction or redefined to the tube direction as the $z$ direction. 

Similarly, we assume that the segregate planar state undergoes symmetry-breaking in the $z$ direction and define the correlation function as 
\begin{equation}
	g_{\perp}(r)=\braket{\rho_{\perp}(z_1)\rho_{\perp}(z_2)}/\rho_{\mathrm{3D}}^2,
\end{equation}
where $\rho_{\perp}(z)=(L_x L_y)^{-1}\int^{L_x}_0\int^{L_y}_0\rho(x,y,z)\mathrm{d}x\mathrm{d}y$ corresponds to the average density distribution along $z$ direction with $L_x$ and $L_y$ being the length of the numerical box along $x$ and $y$ directions, respectively.

\section{IV. Superfluid fraction}
\begin{figure}[!hb]
	\renewcommand\thefigure{S3}
	\centering
	\includegraphics[width=\columnwidth]{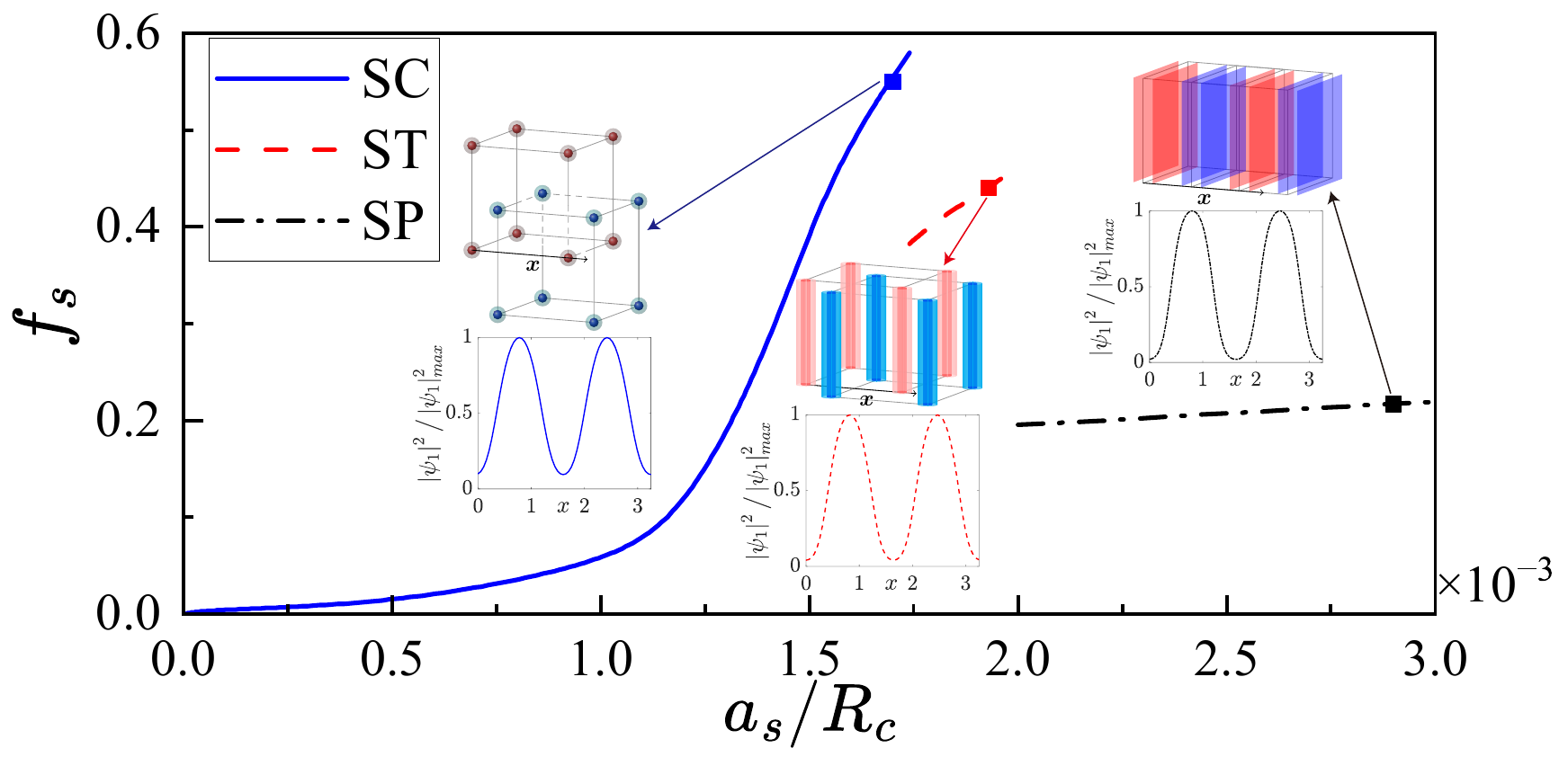}
	\caption{Superfluid fraction $f_s$ of the SC, ST, and SP states along the $x$ direction varies with the contact interaction. Here, we have fixed $R^{(12)}_c/R_c=0.95$, and the other parameters are the same as in Fig. 2 in the main text.} 
	\label{figs3}
\end{figure}

In the main text, we focus on the ground-state structures that exhibit a variety of three-dimensional spatial symmetry breakings, arising from the interplay between the ranges of inter- and intra-component interactions. Due to the superfluid nature of the RdBECs, the phases shown in Fig.~2 also exhibit supersolid properties, revealing a rich spectrum of 3D supersolid states. To confirm this, we examine the superfluid fraction~\cite{PhysRevLett.25.1543,Leggett1998} of representative phases such as the SC, ST, and SP states. As shown in Fig.~\ref{figs3}, the density profiles of these states exhibit finite overlap between neighboring sites—a hallmark of supersolidity. For the SC state, the superfluid fraction gradually goes up with increasing contact interaction. Beyond the critical point of $a_s/R_c=1.8\times 10^{-3}$, the ground state transitions to the ST phase, accompanied by a sudden drop in the superfluid fraction. Upon further increasing the contact interaction past the second critical point at $a_s/R_c=2\times 10^{-3}$, a second drop in the superfluid fraction is observed, signaling the onset of the SP phase. Such a finite but sub-unity superfluid fraction is a distinctive feature of supersolidity~\cite{PhysRevLett.25.1543,Leggett1998,Tanzi2019PRL,Chomaz2019PRX,Bottcher2019PRX}. 

\bibliography{mybib}